\documentclass[usenatbib]{mnras}

\usepackage[T1]{fontenc}

\DeclareRobustCommand{\VAN}[3]{#2}
\let\VANthebibliography\thebibliography
\def\thebibliography{\DeclareRobustCommand{\VAN}[3]{##3}\VANthebibliography}

\usepackage{xcolor}

\usepackage[normalem]{ulem}

\usepackage{graphicx}	
\usepackage{amsmath}	
\pdfoutput=1 
\usepackage{amssymb}	
\usepackage{booktabs}   
\usepackage{color}
\usepackage{txfonts}
\usepackage{natbib}
\usepackage{pdflscape} 
\usepackage{subcaption}
\usepackage{lmodern}
\usepackage{tabularx}

\newcommand{\teff}{T_{\rm{eff}}}

\title[Colour$-\teff$ relations in the SDSS system]{Empirical colour–effective temperature relations in the SDSS system from IRFM temperatures of GALAH and APOGEE stars}

\author[Z. H. Zhou et al.]{
Zenghua~Zhou$^{1,2,5,6}$\thanks{E-mail: \href{zhouzenghua@ynao.ac.cn}{zhouzenghua@ynao.ac.cn}}
Luca~Casagrande$^{2}$,
Xiaobin~Zhang$^{4,6}$,
Jianping~Xiong$^{1,5}$,
Jiao~Li$^{1,5}$,\newauthor
Yanjun Guo$^{1,5}$,
Zhanwen~Han$^{1,3,5,6}$,
Xuefei~Chen$^{1,3,5,6}$\thanks{E-mail: \href{mailto:cxf@ynao.ac.cn}{cxf@ynao.ac.cn}}
\\
\\
$^{1}$Yunnan Astronomical Observatories, Chinese Academy of Sciences, Kunming 650216, China\\
$^{2}$Research School of Astronomy and Astrophysics, Australian National University, Weston Creek ACT 2611, Australia\\
$^{3}$Key Laboratory for the Structure and Evolution of Celestial Objects, Chinese Academy of Sciences, Kunming 650011, China\\
$^{4}$CAS Key Laboratory of Optical Astronomy, National Astronomical Observatories, Chinese Academy of Sciences, Beijing 100101,
China\\
$^{5}$International Centre of Supernovae (ICESUN), Yunnan Key Laboratory of Supernova Research, Kunming 650216, China\\
$^{6}$University of the Chinese Academy of Sciences, Yuquan Road 19, Shijingshan Block, Beijing 100049, China\\
}

\date{Accepted XXX. Received YYY; in original form ZZZ}

\pubyear{\the\year{}}

\begin{document}
\label{firstpage}
\pagerange{\pageref{firstpage}--\pageref{lastpage}}
\maketitle

\begin{abstract}
Reliable estimates of stellar effective temperature ($\teff$) are fundamental to stellar population studies and Galactic astrophysics. However, the majority of stars observed in modern large-scale photometric surveys lack spectroscopic measurements, making empirical colour--$\teff$ relations essential tools. In this work, we present updated empirical colour--$\teff$ calibrations based on Sloan Digital Sky Survey (SDSS) $ugriz$ photometry combined with 2MASS $JHK_{\rm s}$ data. Effective temperatures are determined on a homogeneous InfraRed Flux Method (IRFM) scale using a combined sample of 3902 GALAH and 2535 APOGEE stars with high-quality photometry and well-characterised atmospheric parameters. Using this dataset, we establish empirical relations between $\teff$ and colour indices constructed from SDSS and 2MASS combinations. We provide both colour--metallicity--$\teff$ and colour--$\teff$ relations for dwarfs and giants. The calibrations are derived using low-order polynomial models with iterative $3\sigma$ clipping. Their performance depends on the adopted colour index, with long-baseline colours such as $(g-K_{\rm s})_0$ and $(g-z)_0$ achieving internal precisions of $\sim$30--50~K. Comparisons with previous calibrations show general agreement, with differences attributable to sample selection, photometric zero-points, and functional form. The resulting relations provide a homogeneous and internally consistent framework for estimating $\teff$ from SDSS and 2MASS photometry alone, and are well suited for application to large photometric surveys lacking spectroscopic information.
\end{abstract}




\section{Introduction}
Accurate stellar effective temperatures ($\teff$) are fundamental to a wide range of problems in stellar and Galactic astrophysics, underpinning the determination of nearly all other stellar properties, including radii, luminosities, chemical abundances, ages, and distances \citep{Allende1999A&A, Boyajian2012ApJ, Heiter2015A&A, Teff_ages}. High-resolution spectroscopy enables precise determination of atmospheric parameters with well-characterised uncertainties \citep[e.g.,][]{Valenti2005ApJS, Sousa2008A&A, Bensby2014A&A, Jofre2014A&A, Smiljanic2014A&A, Hinkel2016ApJS}. However, such observations are observationally expensive and remain unavailable for the vast majority of stars targeted by modern large-scale photometric surveys. Direct interferometric measurements are also limited to bright, nearby stars and cannot be extended to the faint or distant populations that are critical for Galactic archaeology. Consequently, empirical colour--effective temperature ($\teff$) relations remain indispensable, providing robust and observationally economical estimates of $\teff$ from broadband photometry alone. These relations enable the homogeneous characterisation of millions of stars in large surveys such as GALAH, Gaia, and 2MASS \citep[e.g.,][]{Casagrande2021MNRAS}.

The use of broadband photometric colours to estimate $\teff$ has a long history, with early calibrations based on optical and near-infrared colours \citep[e.g.][]{Johnson1966ARA&A, Blackwell1977MNRAS, Bessell1979PASP, Blackwell1980A&A, Blackwell1990A&A, Alonso1996A&A, Alonso1999A&AS, Flower1996ApJ, Sekiguchi2000AJ}. Subsequent studies refined these relations using increasingly homogeneous datasets, improved extinction corrections, and physically motivated temperature scales \citep[e.g.][]{Ramirez2005ApJ, González2009A&A, Casagrande2006MNRAS, Casagrande2010A&A, Huang2015MNRAS, Mucciarelli2020RNAAS, Mucciarelli2021A&A, Casagrande2019MNRAS, Casagrande2021MNRAS}. Despite these advances, existing colour--$\teff$ relations are not always mutually consistent, which limits their direct applicability, particularly when combining data from multiple surveys.

Among the available methods for determining $\teff$, the InfraRed Flux Method \citep[IRFM;][]{Blackwell1977MNRAS, Blackwell1979MNRAS, Blackwell1980A&A, Blackwell1994A&A} provides a temperature scale closely tied to the fundamental definition of $\teff$. By comparing observed bolometric fluxes to monochromatic infrared fluxes, the IRFM yields an absolute temperature scale that is only weakly dependent on stellar atmosphere models and relatively insensitive to metallicity ([Fe/H]) and surface gravity ($\log g$). Over the past decades, the IRFM has been widely applied across different spectral types, metallicities, and evolutionary stages \citep{Alonso1996A&A, Ramirez2005ApJ, González2009A&A, Casagrande2006MNRAS, Casagrande2010A&A, Casagrande2021MNRAS}. These studies have benefited from improvements in absolute flux calibration and homogeneous photometric systems, 
yielding temperature scales with internal consistencies at the level of a few tens of Kelvin. Such scales provide an ideal foundation for empirical colour--$\teff$ calibrations.

The Sloan Digital Sky Survey \citep[SDSS;][]{York2000AJ, Fukugita1996AJ, Gunn2006AJ, Doi2010AJ}, combined with near-infrared photometry from the Two Micron All Sky Survey \citep[2MASS;][]{Skrutskie2006AJ}, provides extensive wavelength coverage for millions of stars across the Milky Way. Several colour--$\teff$ relations based on SDSS photometry have been proposed \citep[e.g.][]{Ivezi2008ApJ, Pinsonneault2012ApJS, Huang2015MNRAS}. However, differences in adopted temperature scales, photometric zero-points, extinction treatments, and underlying stellar samples introduce systematic discrepancies between these calibrations. In particular, small zero-point offsets in SDSS photometry \citep{Bohlin2001AJ, Holberg2006AJ, SDSS_DR2, Eisenstein2006ApJS, BBSuzuki2018AJ} propagate directly into IRFM-based $\teff$ estimates, and therefore into the resulting colour--$\teff$ relations. In a companion study \citep{zhou26}, we recalibrated the SDSS $ugriz$ photometric system onto the AB magnitude scale using the IRFM. That work demonstrated the presence of small but non-negligible zero-point offsets, with mild dependencies on stellar parameters. As colour--$\teff$ relations are sensitive to such offsets, these systematics set a floor on the achievable accuracy of photometric temperature estimates.

In this work, we revisit the calibration of colour--$\teff$ relations in the SDSS and 2MASS systems using a homogeneous temperature scale derived by applying the IRFM to corrected SDSS photometry. We establish a self-consistent set of relations for FGK-type stars spanning $4000 \leq T_{\rm eff} \leq 7000\,\mathrm{K}$ and low reddening. Separate calibrations are constructed for dwarfs and giants to account for surface gravity effects. The paper is structured as follows. In Section~\ref{data} we describe the spectroscopic samples, photometric data, and parameter coverage. Section~\ref{method} outlines the methodology used to derive the colour--$\teff$ relations. The results are presented in Section~\ref{result}, with discussion and conclusions in Section~\ref{sec:concl}.

\section{Data}\label{data}
\subsection{Spectroscopic reference samples}
The stellar samples used in this work are drawn from two high-quality spectroscopic surveys: GALAH DR3 \citep[Galactic Archaeology with HERMES;][]{GALAH2015MNRAS, Buder2021MNRAS} and APOGEE \citep[Apache Point Observatory Galactic Evolution Experiment;][]{APOGEEdr172022ApJS}. We adopt the catalogues compiled by \citet{Casagrande2021MNRAS} for GALAH DR3 stars and \citet{Govind2022MNRAS} for APOGEE stars. The latter catalogue was calibrated onto the GALAH DR3 metallicity scale, ensuring greater homogeneity in our sample. To establish a robust reference set for calibrating empirical colour--$\teff$ relations, we retain only stars with high-quality spectroscopic parameters and reliable photometry. This selection results in cleaned samples of 3902 GALAH stars and 2535 APOGEE stars \citep{zhou26}, with further cuts applied in Section \ref{sec:photo}.

\begin{figure}
    \centering
    \includegraphics[width=\columnwidth]{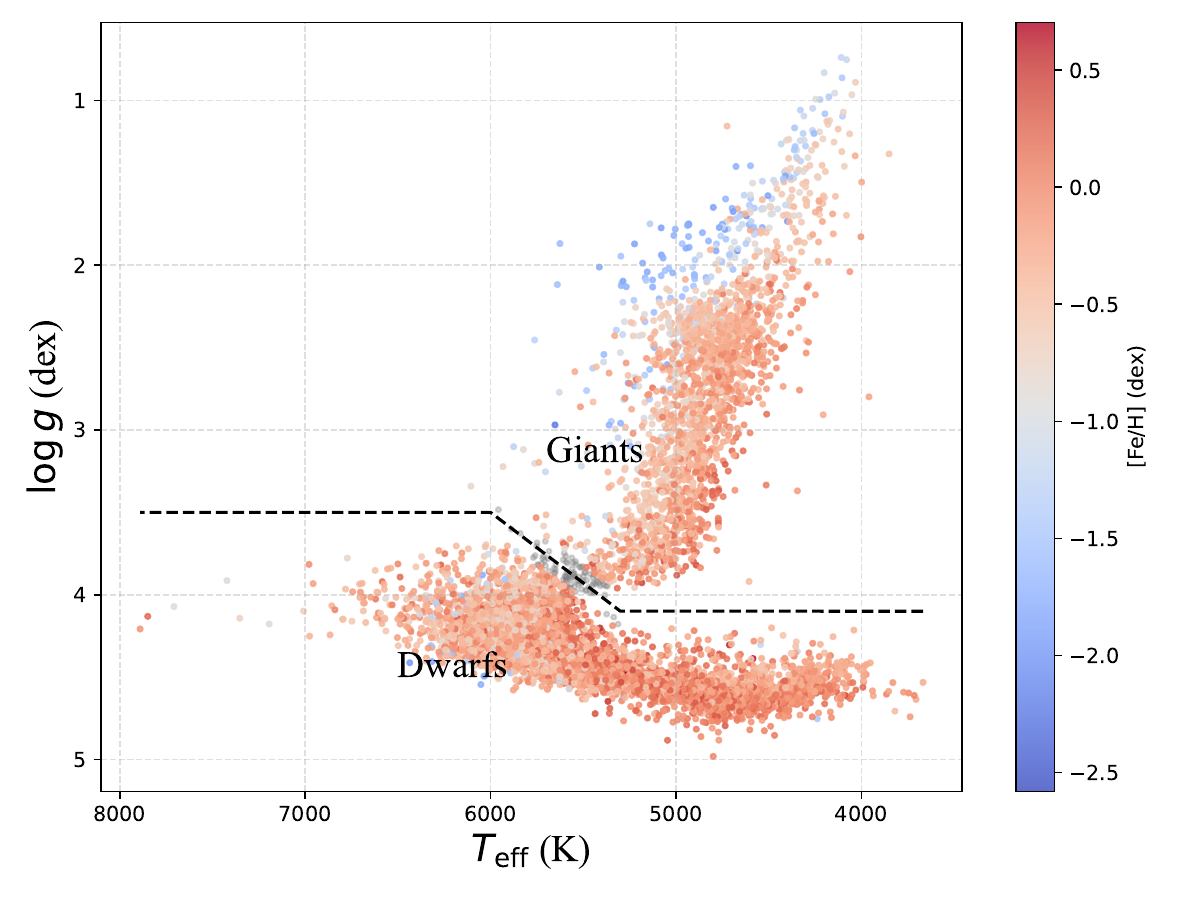}
    \caption{Distribution of the combined GALAH and APOGEE samples in the $T_{\mathrm{eff}}$–$\log g$ plane, colour-coded by metallicity [Fe/H]. The dashed line indicates the adopted temperature-dependent separation between dwarfs and giants. Stars within the buffer region around this boundary (shown in grey) are excluded from the calibration but retained for validation purposes.} 
    \label{fig:parameters}
\end{figure}

The two samples are combined to derive the colour--$\teff$ calibrations presented in this paper. Figure~\ref{fig:parameters} shows the distribution of the combined sample in the $T_{\mathrm{eff}}$–$\log g$ plane, colour-coded by metallicity. The stars span an effective temperature range of approximately $3800 \lesssim \teff \lesssim 8000$~K. Since only a small number of stars fall below 4000~K or above 7000~K and these limits define the range within which our calibrations have been build, and should be used (Section \ref{method}). The GALAH sample is dominated by dwarf and subgiant stars, with $\log g$ primarily between $\sim$3.5 and 4.5, and exhibits a metallicity distribution representative of the solar neighbourhood. In contrast, the APOGEE sample covers a broader range of evolutionary stages, including red giant branch stars ($\log g \sim 2.5$--3.5) and main-sequence or turn-off stars ($\log g \sim 4.3$--4.6). The metallicity and $\teff$ distributions of the two samples overlap substantially, ensuring consistency when combining them for calibration. This combined dataset provides continuous coverage across evolutionary stages and metallicity, which is essential for deriving robust and homogeneous colour--$\teff$ relations. 

\subsection{Photometric data and reddening}\label{sec:photo}

All stars used in this work have photometry from SDSS DR13 $ugriz$ \citep{SDSSDR132017ApJS}, 2MASS $JHK_{\rm s}$ \citep{Skrutskie2006AJ}
and Gaia $BP$ and $RP$ \citep{riello}. 
Corrections to the SDSS photometric zero-points are derived as described in \cite{zhou26} and summarised in Section \ref{sec:summary}, with Gaia photometry used only to establish the reference sample. 
Here it suffices to say our SDSS $ugriz$ zero-point corrections are used only within the IRFM for the purpose of deriving accurate effective temperatures. Importantly, when using the calibrations provided in Section \ref{method} and Appendix \ref{appendix}, the published SDSS photometry should be used as is, with no zero-point corrections applied.
\begin{figure}
\centering
\includegraphics[width=\columnwidth]{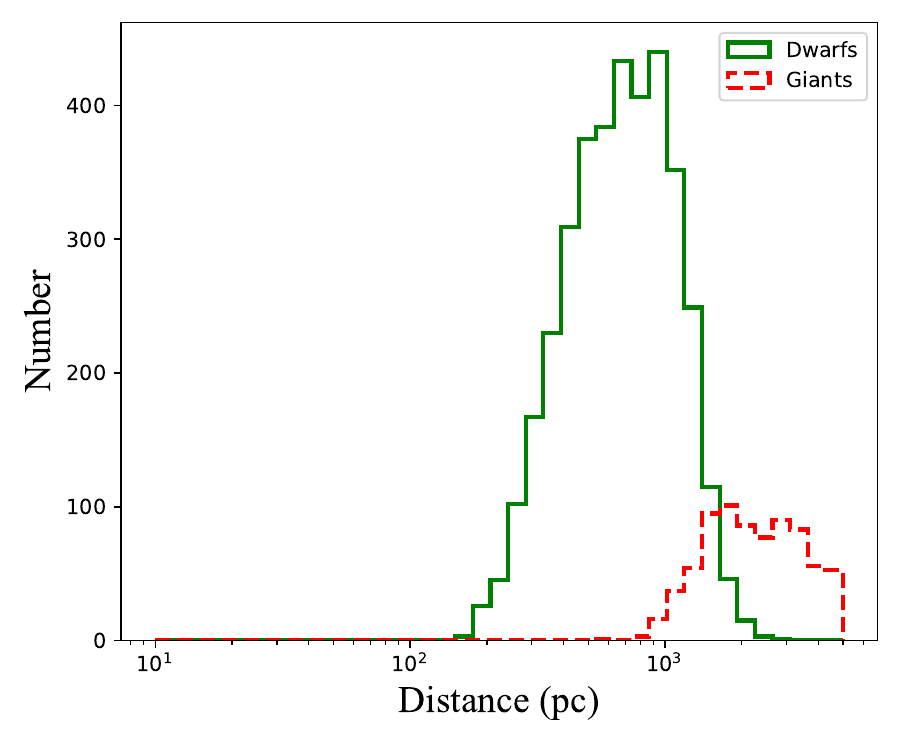}
\caption{
Distance distributions of the dwarf (green solid line) and giant (red dashed line) samples derived from Gaia parallaxes. Dwarfs are predominantly nearby, while giants probe larger distances.
}
\label{fig:distance}
\end{figure}

Interstellar reddening corrections are applied using line-of-sight colour excess values $E(B-V)$ primarily derived from the three-dimensional Bayestar2019 dust map \citep{Green2019ApJ}, which provides distance-dependent extinction estimates based on stellar photometry. For a fraction of stars not covered by this map, we adopt values from \citet{Schlegel1998ApJ}, rescaled as described in \cite{Casagrande2019MNRAS}. To further minimize uncertainties associated with extinction, we retaining only stars with $E(B-V) \leq 0.10$. This requirement reduces the sample sizes to 3375 stars for GALAH and 1600 stars for APOGEE. The removed stars are primarily those affected by stronger interstellar reddening, particularly more distant giants from APOGEE (cf. Fig.~\ref{fig:distance}). Stellar distances are derived by inverting Gaia DR3 parallaxes and used only for the purpose of distance interpolating the Bayestar2019 dust map.

It is also worth to point out that at low reddening values, stars move approximately along the same colour--$\teff$ relation as reddening is varied. Consequently, although the $\teff$ derived for an individual star depends on the adopted $E(B-V)$, the colour--$\teff$ relations themselves remain robust against reddening uncertainties \citep{Casagrande2021MNRAS}.
Our adopted implementation of the IRFM uses the \citet{CCM1989ApJ} extinction law with $R_V = 3.1$ to compute extinction coefficients for each star based on the synthetic spectrum used at each iteration in the IRFM. All photometric data is dereddened prior to constructing the colour indices of our colour--$\teff$ relations.


\subsection{Effective temperature from improved SDSS $ugrz$ zero-points in the IRFM}\label{sec:summary}

Effective temperatures for the GALAH and APOGEE samples are derived using the IRFM, following the methodology of \citet{Casagrande2010A&A, Casagrande2021MNRAS} which combine multiband photometry. In all instances, 2MASS photometry is used in the infrared. In the optical, SDSS $ugriz$ photometry is implemented after correcting its zero-point as described in \cite{zhou26}. 

In brief, the SDSS zero-points are derived by requiring that the IRFM implementation in the SDSS system reproduces the reference temperature scale obtained when using instead Gaia photometry for optical bands in the IRFM. This procedure yields zero-point corrections to place each $ugriz$ band onto the AB system before this photometry (along with 2MASS) is used in the IRFM to derive effective temperatures for each star in our sample. 

\begin{figure}
    \centering
    \includegraphics[width=0.97\columnwidth]{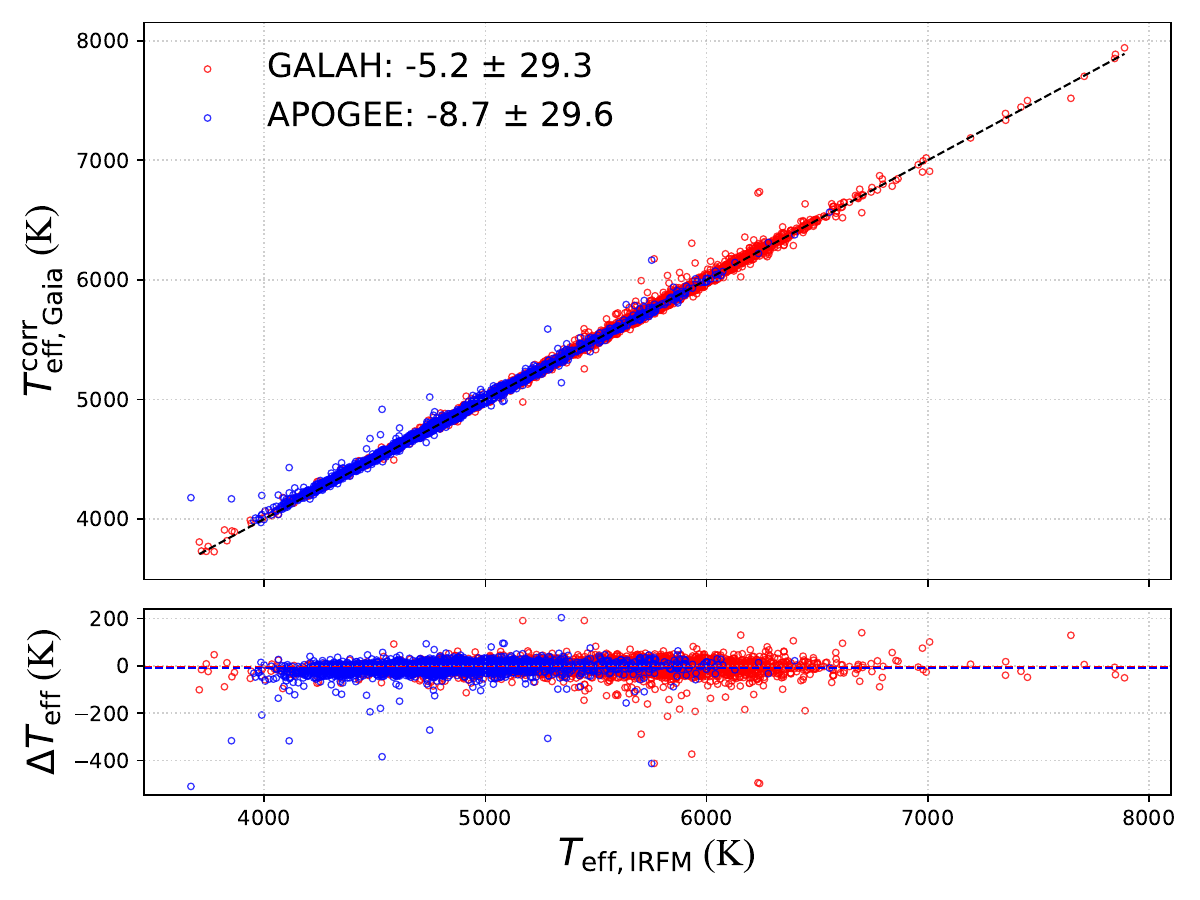}
    \caption{Comparison between effective temperatures derived from SDSS-based IRFM and the Gaia-based IRFM reference scale. Red and blue points represent the GALAH and APOGEE samples, respectively. The mean residual is $\sim -8$~K, with no significant trend as a function of $\teff$.}
    \label{fig:sdssVSgaia}
\end{figure}

Figure~\ref{fig:sdssVSgaia} compares the SDSS-based IRFM $\teff$ with the reference Gaia scale, in the sense of the former minus the latter. The agreement is very good for both GALAH and APOGEE stars, with no significant trends as a function of temperature. The mean offset is small, at the level of approximately $\sim -8$~K. This indicates that the SDSS-based IRFM scale is well calibrated onto the reference scale of \cite{Casagrande2021MNRAS}, which has been validated against both solar twins and interferometric measurements.


\section{Methods}\label{method}
In this section, we describe the methodology adopted to derive empirical colour--$\teff$ relations for the SDSS photometric system. Our approach calibrates $\teff$ as a function of photometric colour indices using stars with reliable temperatures derived implementing SDSS and 2MASS photometry in the IRFM. The calibration is based on samples of nearly 5000 stars drawn from the GALAH and APOGEE surveys, which provide precise stellar parameters, in particular metallicities, allowing us to explicitly account for metallicity effects in the colour--$\teff$ relations. 

Low-order polynomial models are adopted to describe the relations, providing sufficient flexibility to capture non-linear trends while avoiding over-fitting (Section \ref{sec:ct}). To account for structural differences between stellar populations, all relations are derived separately for dwarfs and giants (Section \ref{sec:dg}). 

For those interested in using our calibrations those are available in Table \ref{tab:dwarfs_coeffs} and \ref{tab:giants_coeffs}, with pure colour relations also discussed in Appendix \ref{appendix}. Importantly, when using these calibrations, SDSS $ugriz$ photometry should be used as is, without applying any zero-points correction (see discussion in Section \ref{sec:photo}). 


\subsection{Colour indices and functional form}\label{sec:ct}
We construct several colour indices by combining SDSS $ugriz$ and 2MASS $JHK_{\rm s}$ photometry, thus spanning from the near-ultraviolet to the near-infrared. These indices probe different regions of the stellar spectral energy distribution and provide complementary sensitivity to $\teff$.

Adjacent-band colours, such as $(g-r)_0$ and $(r-i)_0$, primarily trace local spectral slopes and show moderate sensitivity to $\teff$. In contrast, colours spanning a broader wavelength baseline, such as $(g-i)_0$ and combinations involving near-infrared bands, sample a larger fraction of the spectral energy distribution and therefore exhibit a stronger correlation with $\teff$.
The full set of colour indices analysed in this work is reported in the corresponding tables. 

The colour--$\teff$ relations are fitted using polynomial models in the colour index $X$, with optional metallicity-dependent terms. Following standard practice \citep[e.g.,][]{Alonso1996A&A, Alonso1999A&AS, Ramirez2005ApJ, Casagrande2006MNRAS, González2009A&A, Casagrande2010A&A, Huang2015MNRAS, Casagrande2021MNRAS}, we adopt the functional form:

\begin{equation}
\begin{split}
T_{\rm eff} = & \; a_0 + a_1 X + a_2 X^2 + a_3 X^3 + a_4\,\mathrm{[Fe/H]} + a_5 X\,\mathrm{[Fe/H]} \\
& + a_6\,\mathrm{[Fe/H]}^2,
\end{split}
\label{eq:Teff_}
\end{equation}
where $X$ is the dereddened colour index, [Fe/H] stellar metallicity and $a_i$ are fitted coefficients. While in the following Sections we keep using the term colour--$\teff$ relations, those always include a metallicity dependence as per Eq.~\ref{eq:Teff_}. For practical purposes there might be instances where it is useful to have pure colour relations, and we provide those in Appendix \ref{appendix}.

Many previous studies express colour--temperature relations in terms of the inverse temperature parameter $\theta = 5040 / T_{\rm eff}$, which can help linearise the functional dependence on colour. In this work, we instead fit $T_{\rm eff}$ directly. This choice does not affect the physical content of the calibration, as the two formulations are mathematically equivalent under a nonlinear transformation. It also facilitates a more direct physical interpretation of residuals and uncertainties in units of Kelvin. We verified that adopting $T_{\rm eff}$ instead of $\theta$ leads to consistent calibrations within the uncertainties, and fitting in $\theta$ does not significantly reduce the scatter for our sample and colour combinations. Fitting $T_{\rm eff}$ directly also has practical advantages, as it allows temperatures to be estimated without the need for inversion, thereby simplifying application and avoiding the propagation of additional numerical uncertainties. We find that low-order polynomial models in $T_{\rm eff}$ provide an equally good description of the data over the adopted parameter range, without introducing additional transformations.

The polynomial order is selected individually for each colour index based on empirical performance. Third-order terms are used for colours exhibiting stronger curvature (e.g. $(u-g)_0$), second-order polynomials for most indices, and first-order relations for weakly sensitive colours such as $(J-H)_0$. Higher-order terms are avoided to minimise over-fitting and ensure smooth behaviour across the calibration range.

To ensure robust fits and mitigate edge effects \citep{Casagrande2021MNRAS}, the valid colour range for each relation is defined using percentile-based clipping. Only stars within the 1st to 98th percentiles of the colour distribution are retained. This suppresses regions where photometric uncertainties, intrinsic spectral curvature, and parameter correlations degrade the fit.

The coefficients are derived using iterative $3\sigma$ clipping in $\teff$ to remove outliers. The resulting relations are valid only within the adopted colour ranges, which are reported for each index. Extrapolation beyond these ranges is not recommended, as empirical colour--$\teff$ relations are known to become unstable near the boundaries \citep[e.g.,][]{Huang2015MNRAS, Casagrande2021MNRAS}.

\subsection{Dwarfs vs.~giants separation and limits of applicability}\label{sec:dg}

To account for surface gravity effects, we derive independent colour--$\teff$ relations for dwarfs and giants. The two populations are separated using a temperature-dependent boundary in the $T_{\rm eff}$–$\log g$ plane using a piecewise function (see Figure ~\ref{fig:parameters}):
\begin{itemize}
    \item For $T_{\rm eff} \leq 5300\,\mathrm{K}$, $\log g = 4.1$;
    \item For $T_{\rm eff} \geq 6000\,\mathrm{K}$, $\log g = 3.5$;
    \item For $5300 < T_{\rm eff} < 6000\,\mathrm{K}$, the boundary is linearly interpolated.
\end{itemize}
To reduce cross-contamination between the two populations, a buffer region of width $\Delta \log g = 0.10$ dex is introduced around this boundary. Stars are classified as dwarfs or giants only if they lie outside this buffer; stars within this buffer (hereafter, the transition region) are excluded from the fitting procedure. A total of 138 stars lie in this region. 

These transition-region stars are retained for validation purposes, allowing us to assess the consistency between the dwarf and giant calibrations (see Section~\ref{result}). In particular, we verify that the predicted temperatures agree within $\sim$10--30~K for long-baseline colours.

In addition, stars with $T_{\rm eff} > 6200\,\mathrm{K}$ and $\log g < 3.7$ were also removed to avoid contamination from a handful of horizontal branch stars. Overall, our calibrations can be applied to stars in the effective temperature range 4000~K to 7000~K, depending on their evolutionary status. More informative colour limits are provided for dwarfs and giants in their respective tables.

\begin{figure}
\centering
\includegraphics[width=\columnwidth]{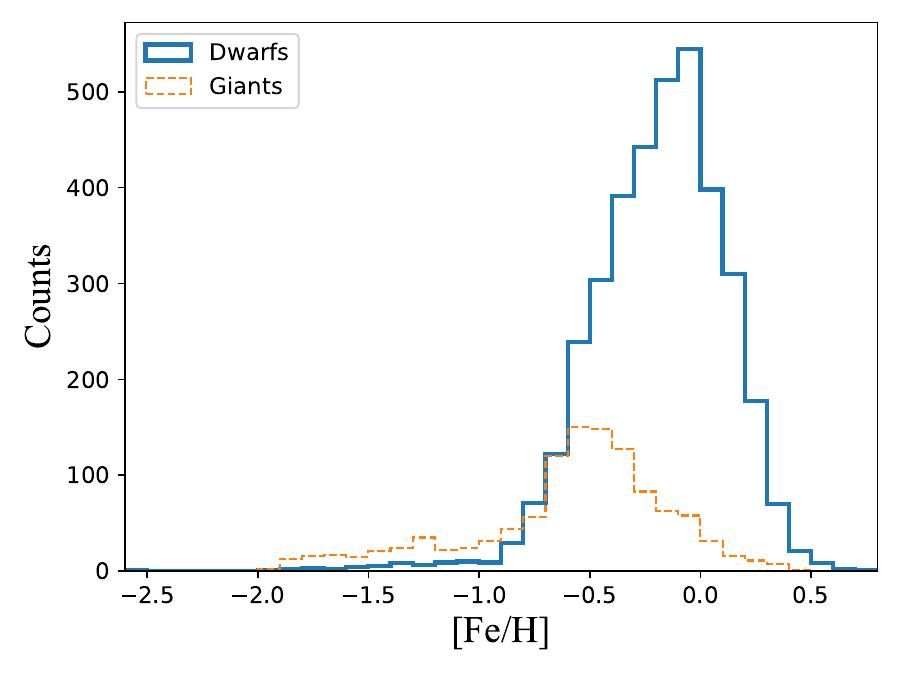}
\caption{
Metallicity distribution functions for the dwarf (blue solid line) and giant (orange dashed line) samples used in the calibration.
}
\label{fig:mdf}
\end{figure}

The metallicity distribution function of the calibration sample is shown in Figure~\ref{fig:mdf}. The combined sample is dominated by stars with near-solar metallicities, with the majority lying in the range $-1.0 \lesssim {\rm [Fe/H]} \lesssim 0.5$. The dwarf sample is strongly concentrated around solar metallicity and contains relatively few metal-poor stars. In contrast, the giant sample peaks at sub-solar metallicities, consistent with the fact that giant stars span a broader range of distances.

This imbalance has implications for the calibration. Although the formal metallicity range covered by our calibrations is broad, the dwarf relations are only weakly constrained at [Fe/H]$ \lesssim-1$. The giant calibrations benefit from a larger fraction of metal-poor stars, but they likewise become progressively less constrained towards lower metallicities. Since the dependence of colour on metallicity weakens at low [Fe/H], we recommend adopting metallicity floors of [Fe/H]$=-1$ for dwarfs and [Fe/H] $=-2$ for giants when applying the calibrations, even for stars with lower measured metallicities.

\section{Calibration performance}\label{result}

\begin{figure*}
    \centering
    \includegraphics[width=0.32\textwidth]{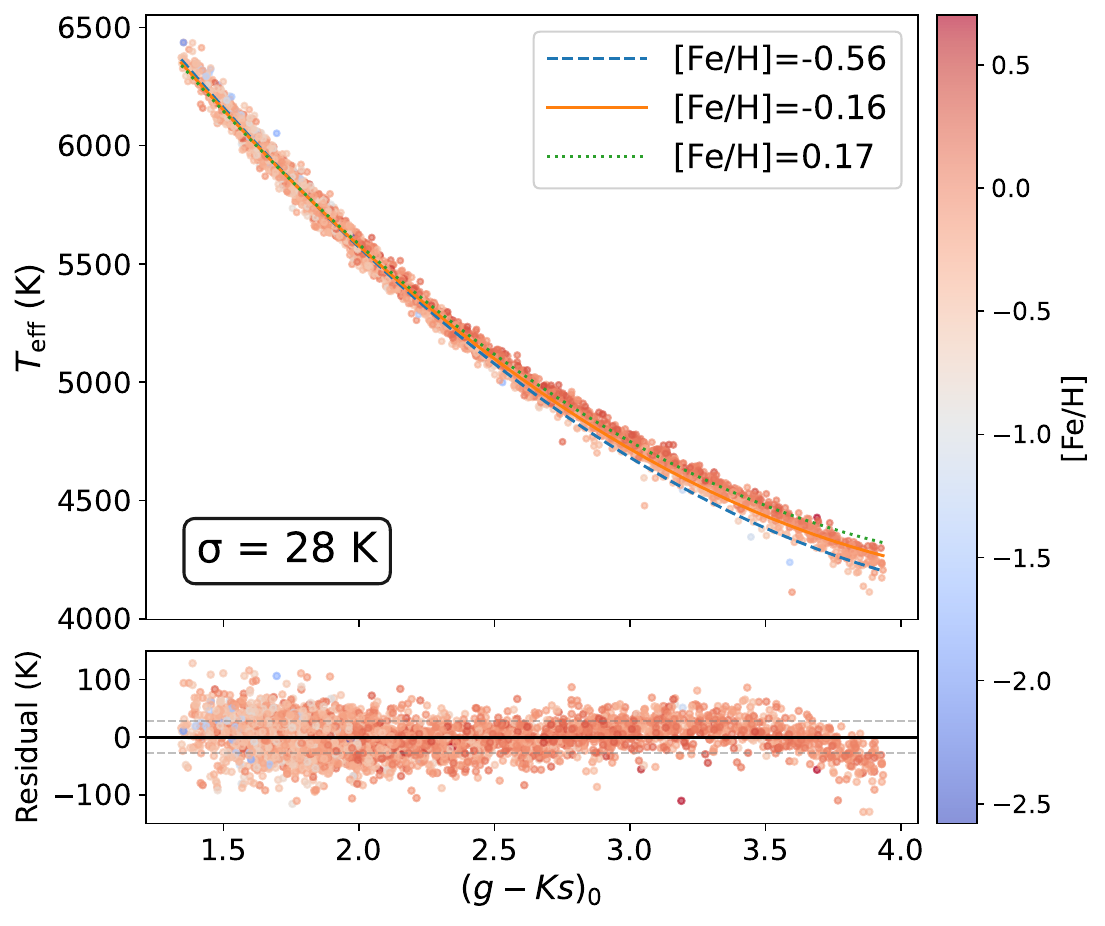}
    \includegraphics[width=0.32\textwidth]{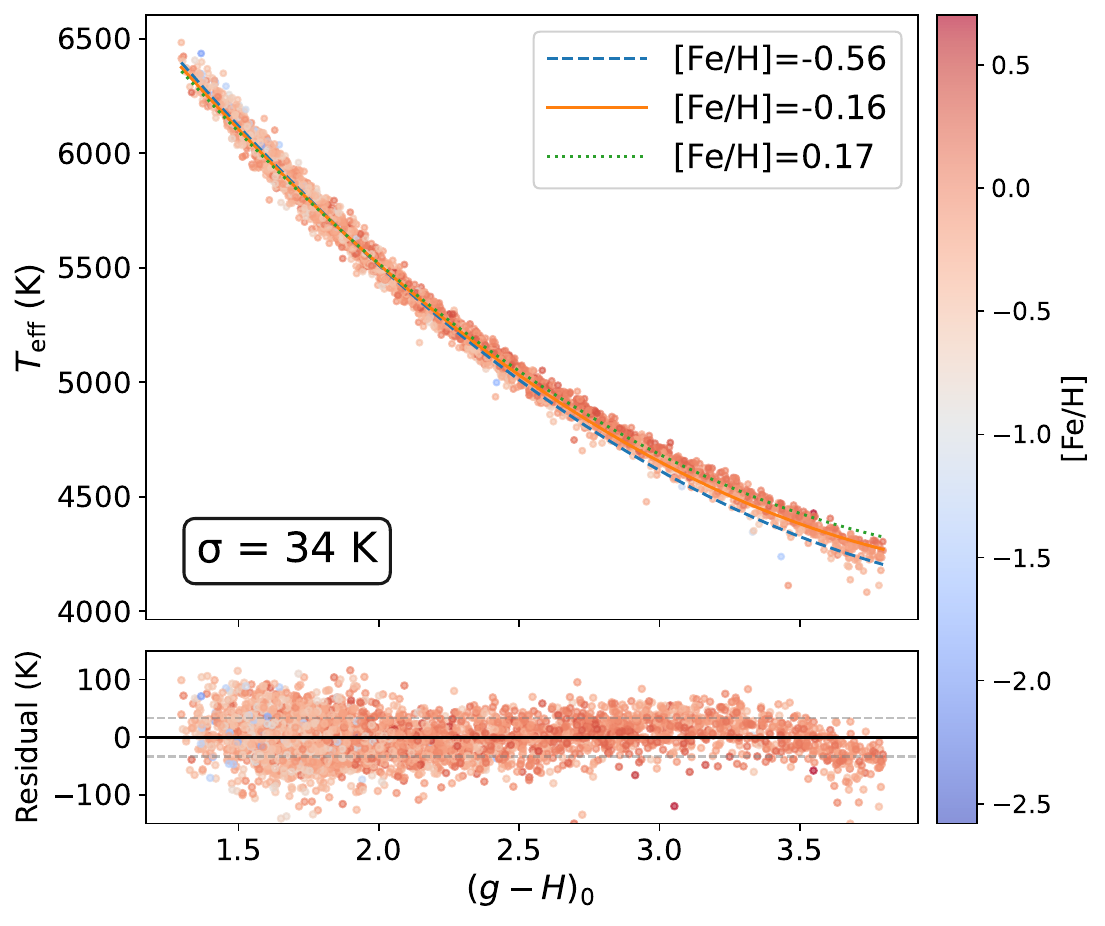}
    \includegraphics[width=0.32\textwidth]{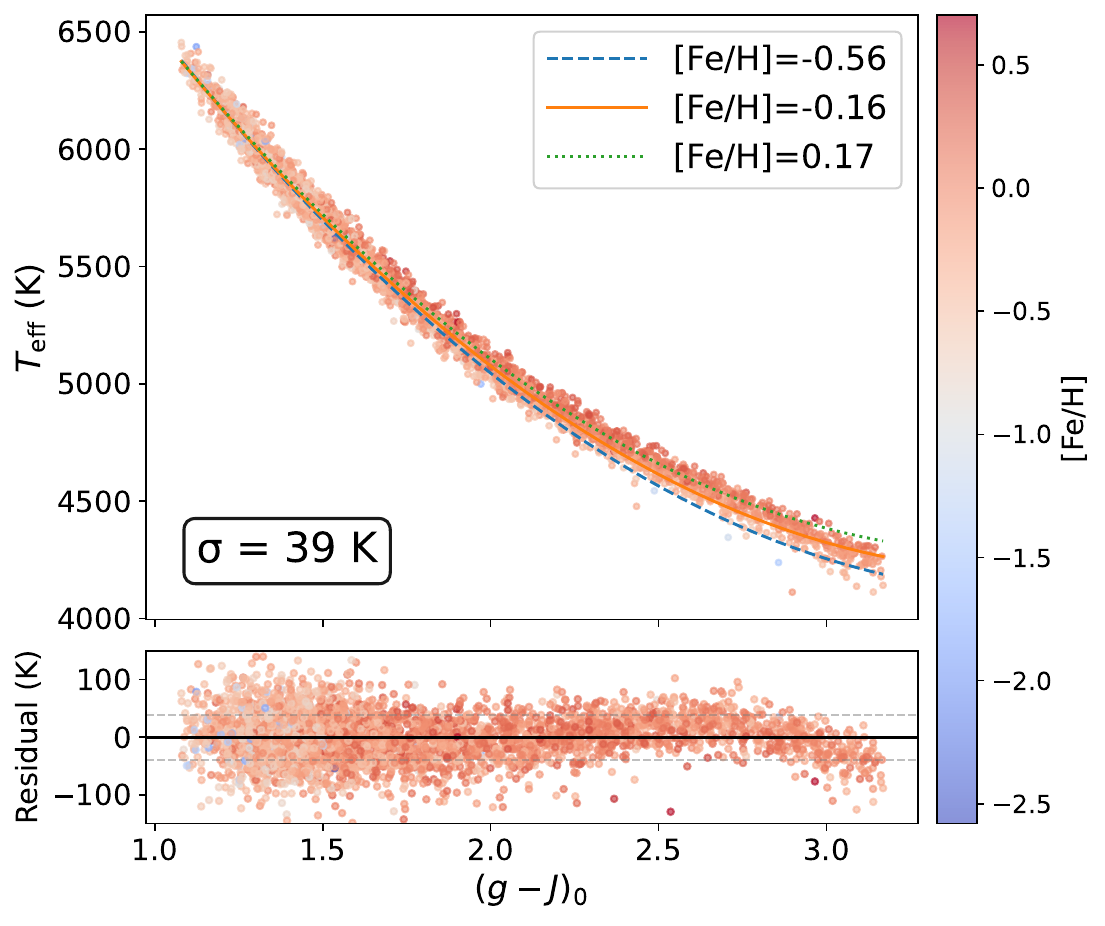}

    \includegraphics[width=0.32\textwidth]{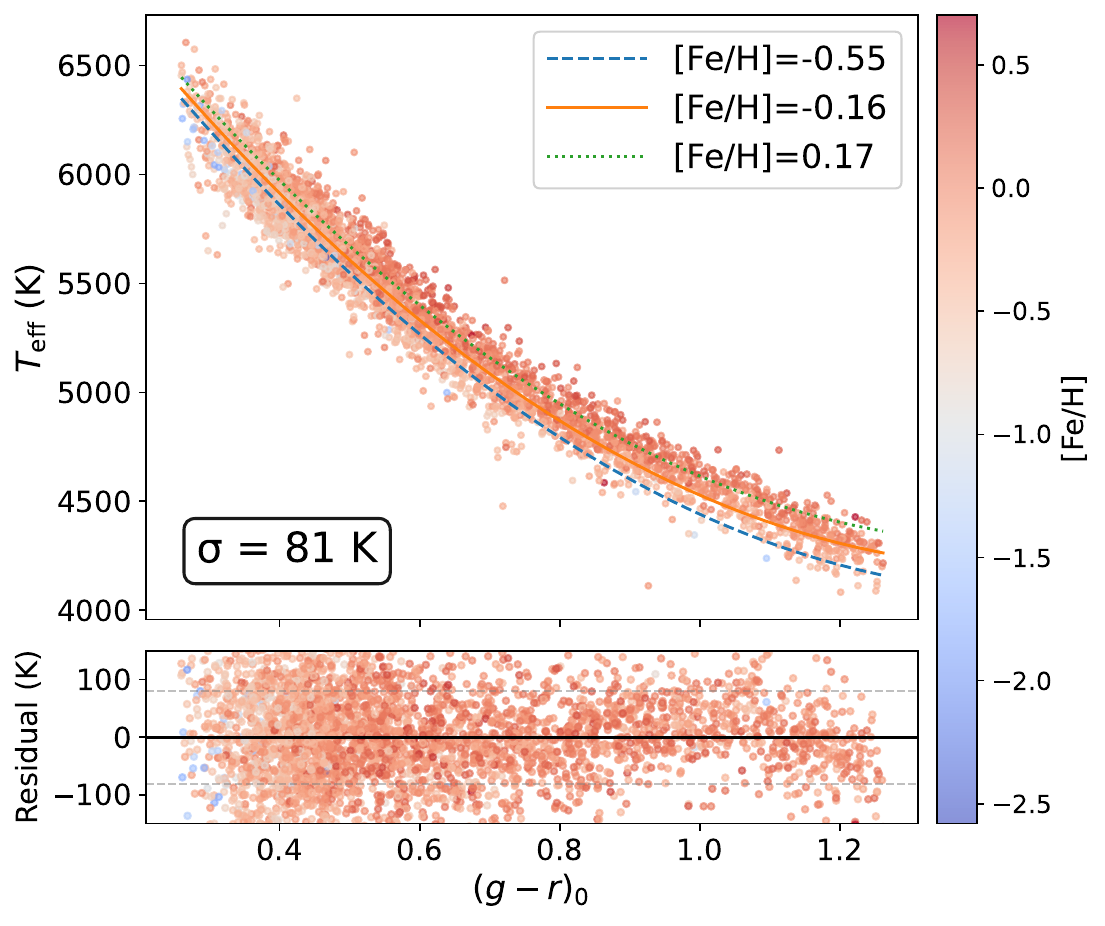}
    \includegraphics[width=0.32\textwidth]{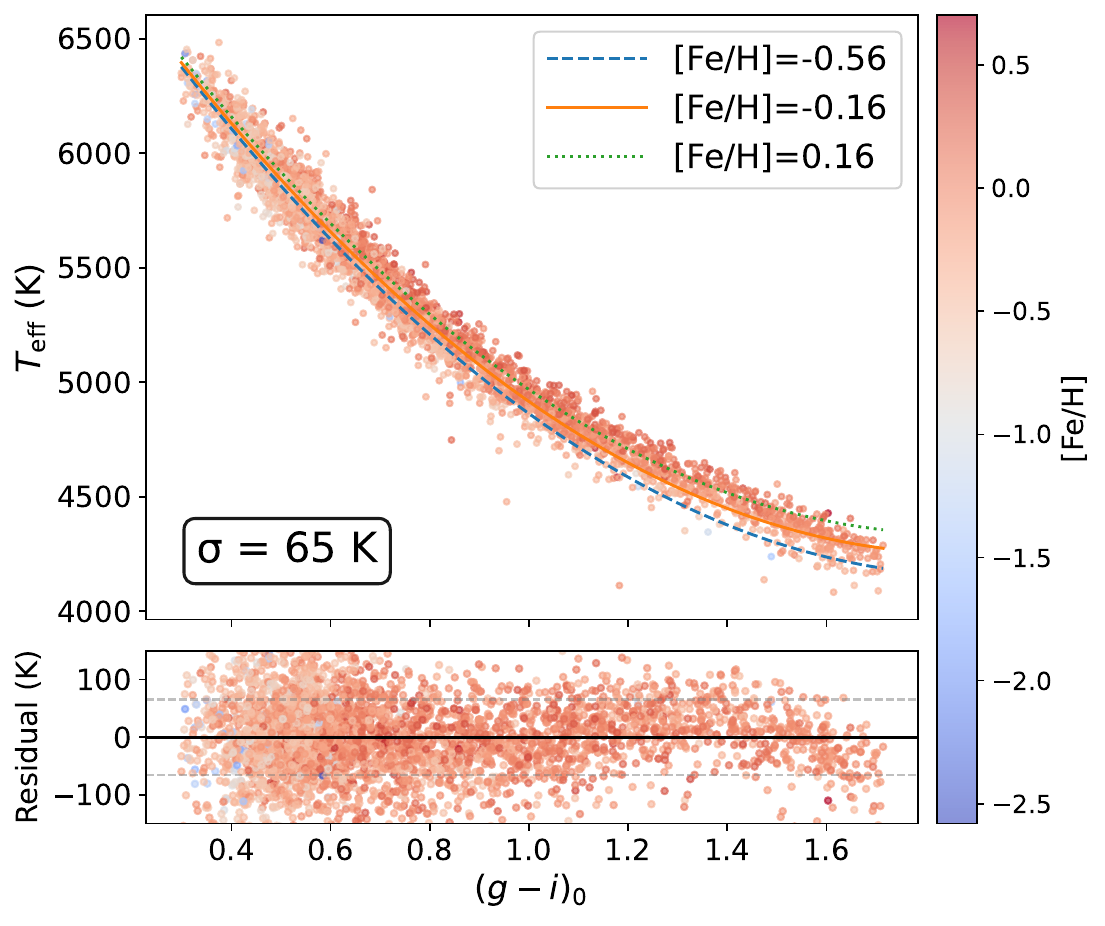}
    \includegraphics[width=0.32\textwidth]{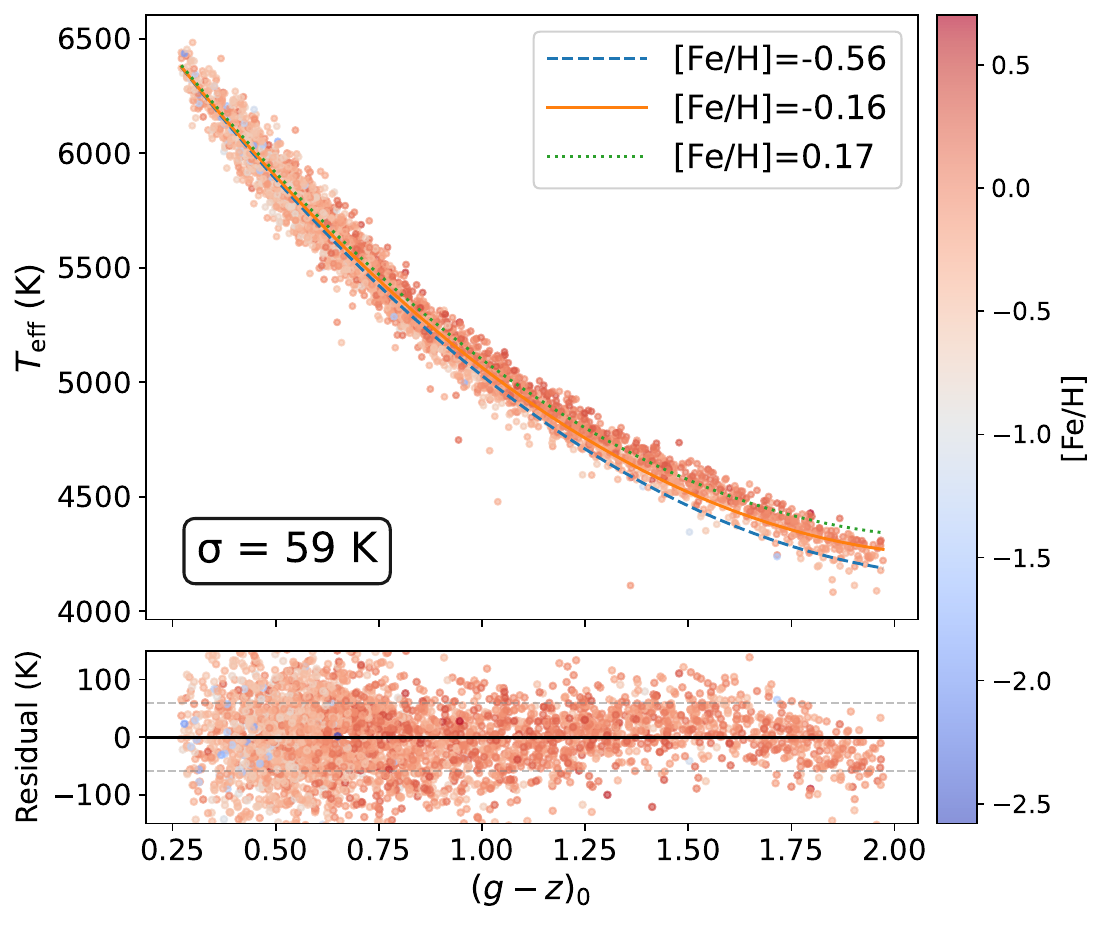}
\caption{Colour-metallicity--$\teff$ relations for dwarfs in a few selected long-baseline (top panels) and optical (bottom) indices. Points are colour-coded by metallicity [Fe/H]. Continuous curves correspond to the best-fitting polynomial relations evaluated at fixed percentiles (10, 50 and 90) of the sample metallicity distribution. Residual are defined as $\Delta T_{\rm eff} = T_{\rm eff}^{\rm IRFM} - T_{\rm eff}^{\rm cal}$. Dashed horizontal lines show the $\pm$RMS scatter of the fit, with standard deviation indicated.
}
\label{fig:example_dwarf_relations}
\end{figure*}

\begin{figure*}
    \centering
    \includegraphics[width=0.32\textwidth]{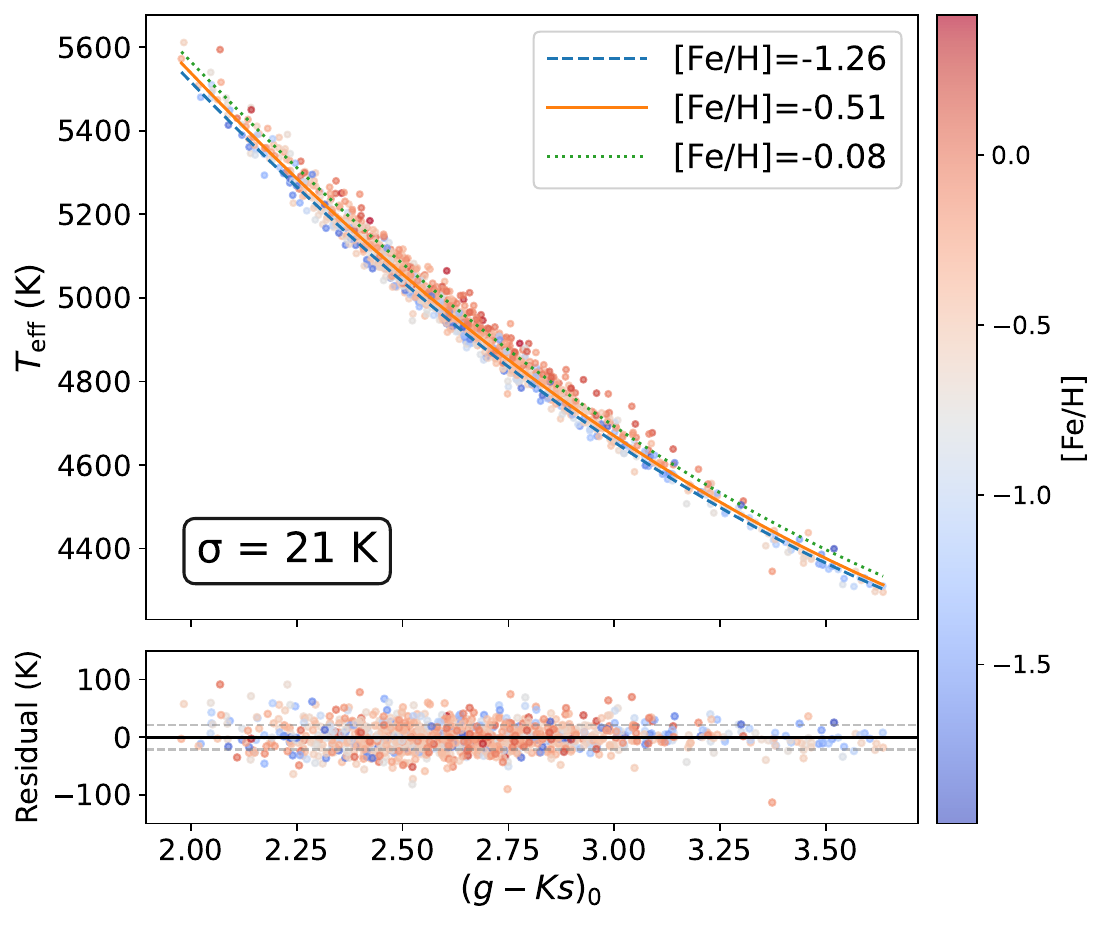}
    \includegraphics[width=0.32\textwidth]{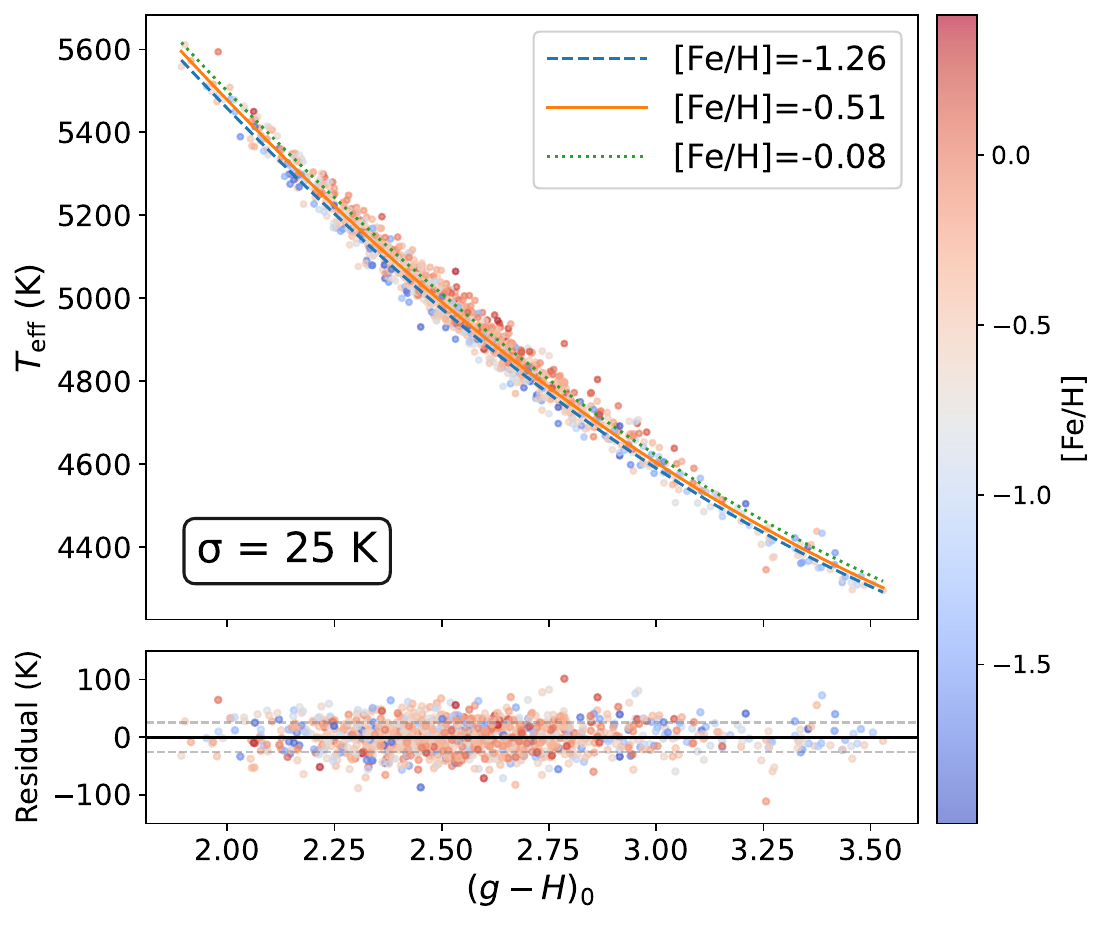}
    \includegraphics[width=0.32\textwidth]{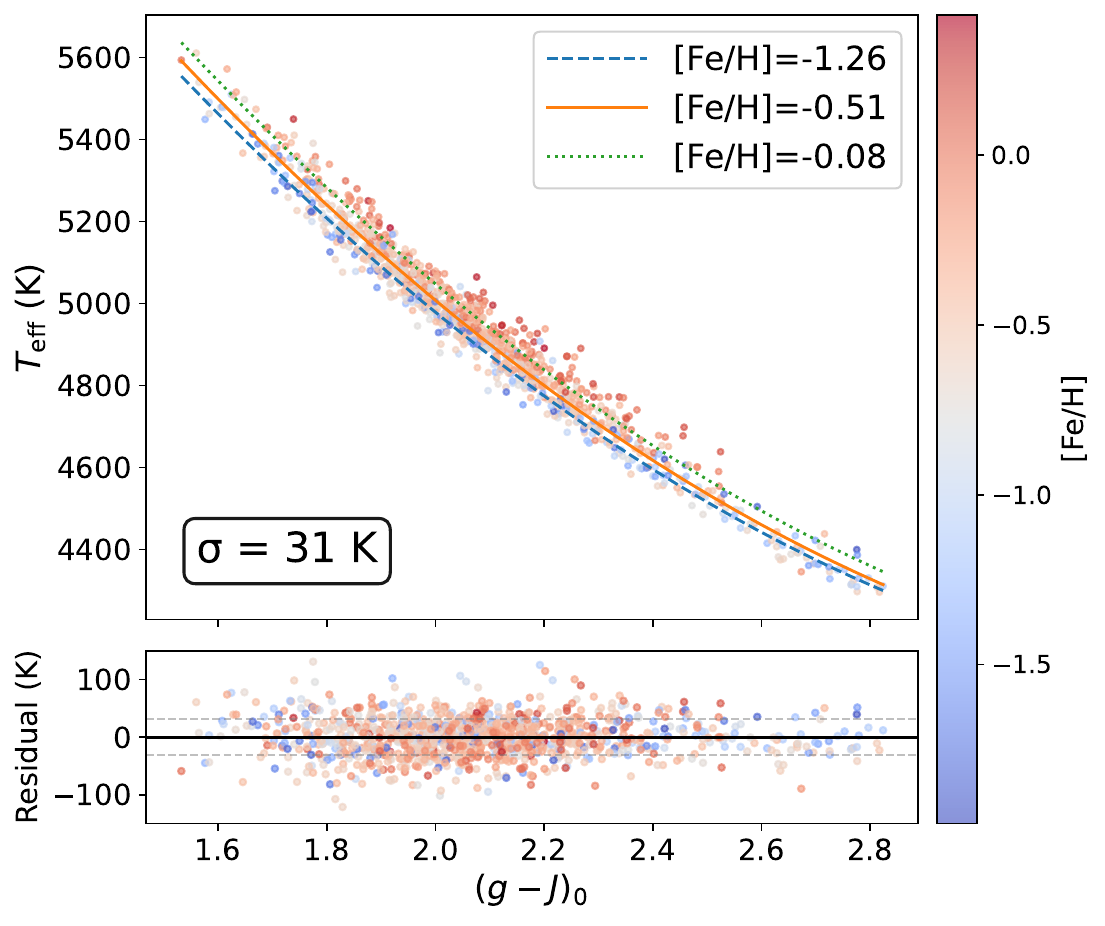}

    \includegraphics[width=0.32\textwidth]{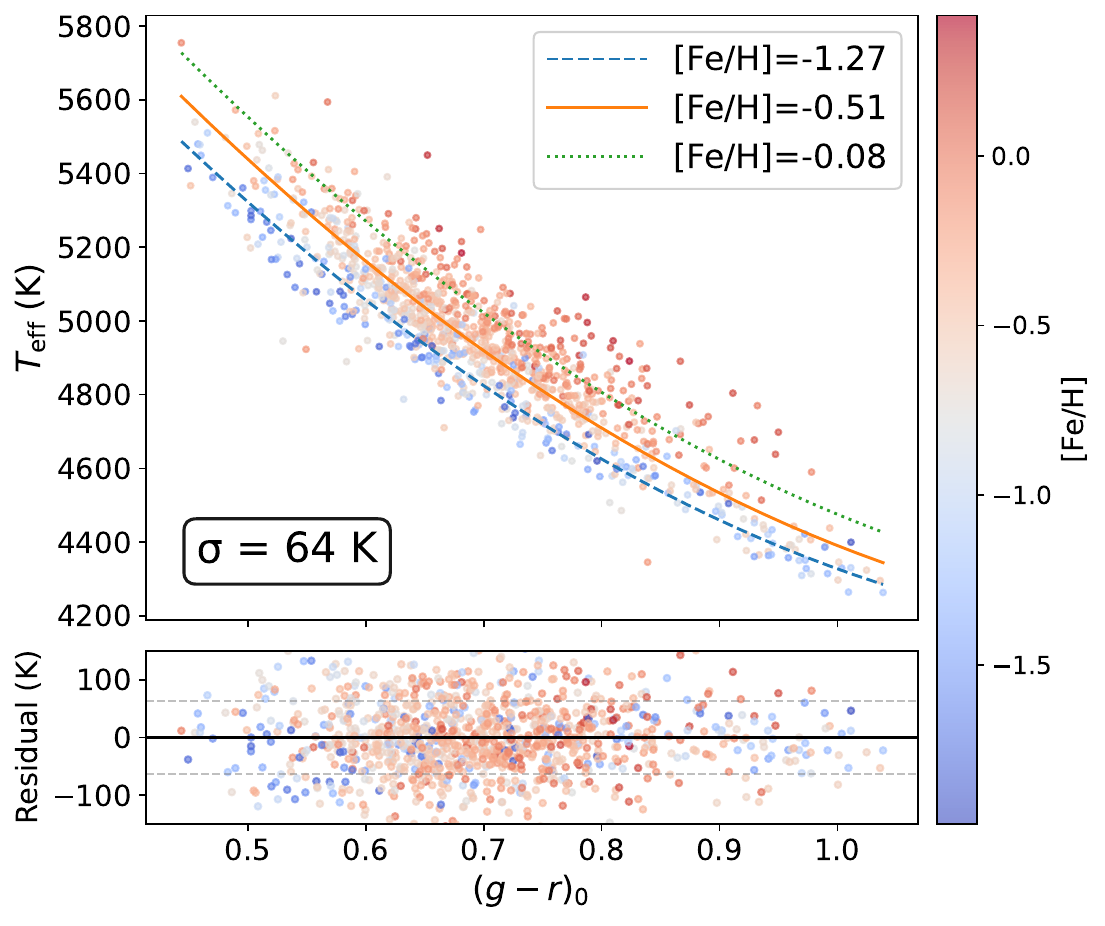}
    \includegraphics[width=0.32\textwidth]{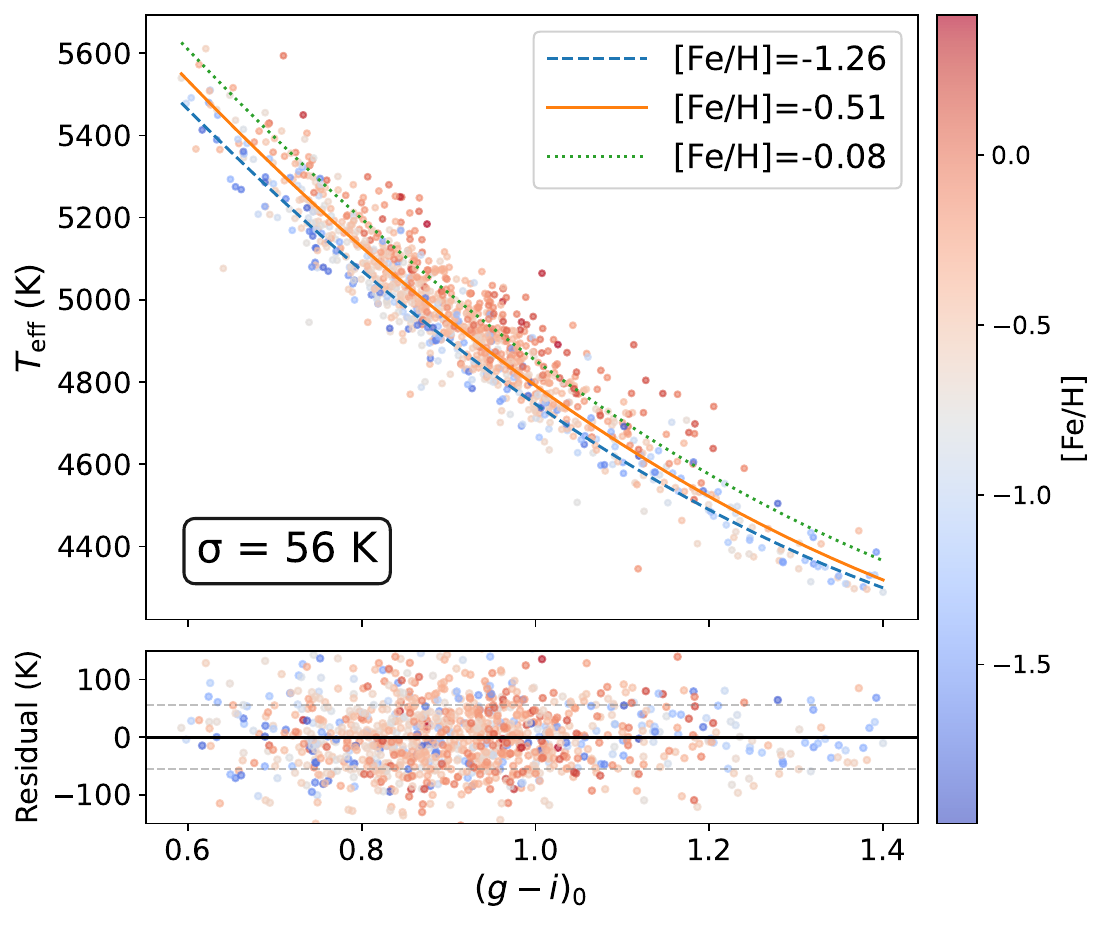}
    \includegraphics[width=0.32\textwidth]{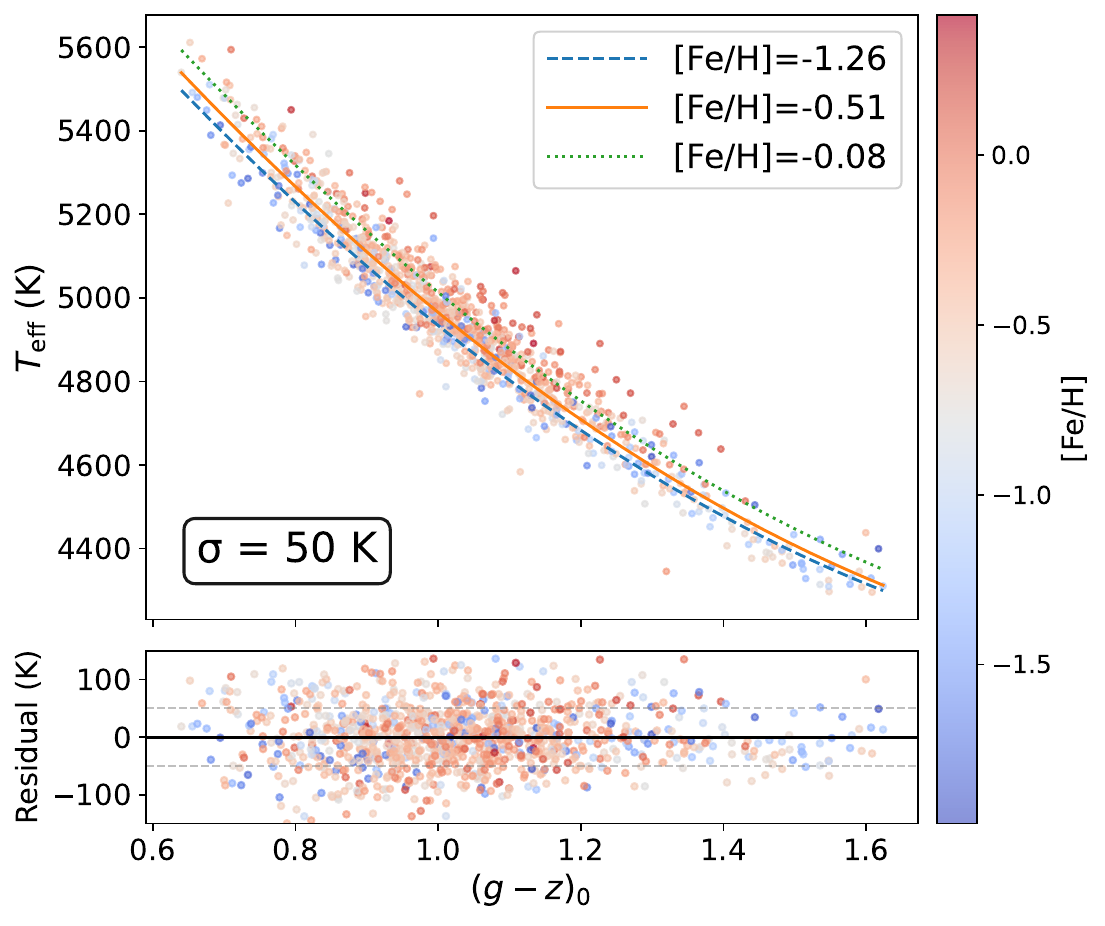}
\caption{Same as Figure~\ref{fig:example_dwarf_relations}, but for the giants.}
\label{fig:example_giant_relations}
\end{figure*}

The derived colour--$\teff$ relations are shown in Figures~\ref{fig:example_dwarf_relations} and \ref{fig:example_giant_relations} for representative dwarf and giant calibrations. In all cases, $\teff$ varies smoothly and monotonically with colour, with metallicity producing systematic shifts that are strongest for optical indices. The adopted polynomial relations reproduce the observed trends well across the calibrated parameter range, with residuals centred close to zero and no significant systematic structure.

The precision of the relations depends strongly on wavelength baseline. Long-baseline optical--infrared colours, such as $(g-K_{\rm s})_0$, $(g-H)_0$, and $(g-J)_0$, provide the tightest constraints on $\teff$, with typical RMS scatter of $\sim$20--40~K. Intermediate optical colours show moderately larger dispersion, while narrow-baseline optical or near-infrared indices exhibit significantly poorer performance. This behaviour reflects the stronger temperature sensitivity and reduced degeneracy with metallicity of broad wavelength-baseline colours.

\begin{table*}
\centering
\small
\setlength{\tabcolsep}{5pt} 
\caption{Polynomial coefficients of our colour--$\teff$ calibrations for dwarf stars. For each colour index $X$, the colour and metallicity range of applicability is given. The adopted functional form is from Equation \ref{eq:Teff_}.}
\label{tab:dwarfs_coeffs}
\begin{tabular}{lcccccccccc}
\hline
\hline
Colour & $a_0$ & $a_1$ & $a_2$ & $a_3$ & $a_4$ & $a_5$ & $a_6$ & Colour range & [Fe/H] range & RMS (K) \\
\hline
$(u-g)_0$ & 10572.151 & -6860.737 & 3079.906 & -538.132 & 753.626 & 120.218 & -114.957 & [0.862, 2.505] & [-2.58, 0.67] & 94.0 \\
$(g-r)_0$ & 7456.020 & -4390.808 & 1503.854 & -- & 122.447 & 65.020 & 145.987 & [0.259, 1.262] & [-2.58, 0.70] & 80.8 \\
$(r-i)_0$ & 6526.687 & -8796.256 & 8210.621 & -- & -281.747 & -44.303 & 832.799 & [0.022, 0.458] & [-2.58, 0.70] & 142.1 \\
$(i-z)_0$ & 5809.473 & -7643.879 & 4388.701 & -- & -288.376 & -51.763 & 854.833 & [-0.057, 0.265] & [-2.58, 0.70] & 192.3 \\
$(z-J)_0$ & 11038.791 & -6830.142 & 970.012 & -- & -479.978 & -2.653 & 427.667 & [0.784, 1.211] & [-2.58, 0.70] & 141.2 \\
$(J-H)_0$& 7111.688 & -4462.782 & -- & -- & -418.936 & -77.127 & 514.397 & [0.201, 0.651] & [-2.58, 0.70] & 149.6 \\
$(g-i)_0$ & 7288.334 & -3205.621 & 857.697 & -- & 38.627 & 39.321 & 122.698 & [0.297, 1.715] & [-2.58, 0.70] & 65.4 \\
$(g-z)_0$ & 7019.751 & -2517.347 & 578.465 & -- & -12.148 & 23.829 & 119.167 & [0.270, 1.974] & [-2.58, 0.70] & 59.0 \\
$(g-J)_0$ & 8613.446 & -2445.075 & 341.623 & -- & -102.096 & 18.962 & 96.294 & [1.071, 3.177] & [-2.58, 0.70] & 39.0 \\
$(g-H)_0$ & 8483.036 & -1907.263 & 211.938 & -- & -163.681 & 1.454 & 86.964 & [1.295, 3.799] & [-2.58, 0.70] & 33.6 \\
$(g-K_{\rm s})_0$ & 8436.528 & -1816.477 & 194.027 & -- & -132.061 & 9.026 & 76.260 & [1.340, 3.936] & [-2.58, 0.70] & 28.2 \\
\hline
\end{tabular}
\end{table*}

\begin{table*}
\centering
\small
\setlength{\tabcolsep}{5pt} 
\caption{Same as Table~\ref{tab:dwarfs_coeffs}, but for giants.}
\label{tab:giants_coeffs}
\begin{tabular}{lcccccccccc}
\hline
\hline
Colour & $a_0$ & $a_1$ & $a_2$ & $a_3$ & $a_4$ & $a_5$ & $a_6$ & Colour range & [Fe/H] range & RMS (K) \\
\hline
$(u-g)_0$ & 8253.675 & -2819.147 & 697.427 & -64.679 & 883.620 & 140.270 & -198.408 & [1.044, 2.546] & [-1.97, 0.41] & 60.0 \\
$(g-r)_0$ & 7492.730 & -4664.907 & 1666.051 & -- & 391.611 & 95.723 & -138.889 & [0.438, 1.040] & [-1.97, 0.41] & 63.8 \\
$(r-i)_0$ & 5924.461 & -5280.955 & 2927.160 & -- & -185.514 & 18.471 & 674.538 & [0.115, 0.376] & [-1.97, 0.41] & 119.6 \\
$(i-z)_0$ & 5340.842 & -3293.307 & -2639.524 & -- & -167.812 & 16.028 & 1325.803 & [0.014, 0.233] & [-1.97, 0.41] & 140.2 \\
$(z-J)_0$ & 7905.306 & -2544.078 & -258.719 & -- & -624.111 & 1.928 & 617.911 & [0.904, 1.222] & [-1.97, 0.41] & 95.1 \\
$(J-H)_0$& 6418.772 & -3064.196 & -- & -- & -297.114 & -71.734 & 301.840 & [0.322, 0.699] & [-1.97, 0.41] & 105.3 \\
$(g-i)_0$ & 7283.563 & -3270.802 & 853.212 & -- & 272.238 & 70.814 & -87.674 & [0.563, 1.401] & [-1.97, 0.41] & 55.9 \\
$(g-z)_0$ & 7006.964 & -2550.451 & 567.233 & -- & 184.234 & 58.889 & -38.251 & [0.590, 1.628] & [-1.97, 0.41] & 50.6 \\
$(g-J)_0$ & 8551.862 & -2381.188 & 316.753 & -- & 166.059 & 46.369 & -22.943 & [1.487, 2.824] & [-1.97, 0.41] & 31.2 \\
$(g-H)_0$ & 8434.309 & -1858.361 & 196.234 & -- & 78.286 & 18.948 & -8.810 & [1.860, 3.531] & [-1.97, 0.41] & 25.5 \\
$(g-K_{\rm s})_0$ & 8442.248 & -1811.319 & 187.729 & -- & 97.919 & 28.773 & -9.184 & [1.927, 3.636] & [-1.97, 0.41] & 21.5 \\
\hline
\end{tabular}
\end{table*}


\subsection{The role of metallicity}

\begin{figure*}
\centering
\includegraphics[width=0.33\textwidth]{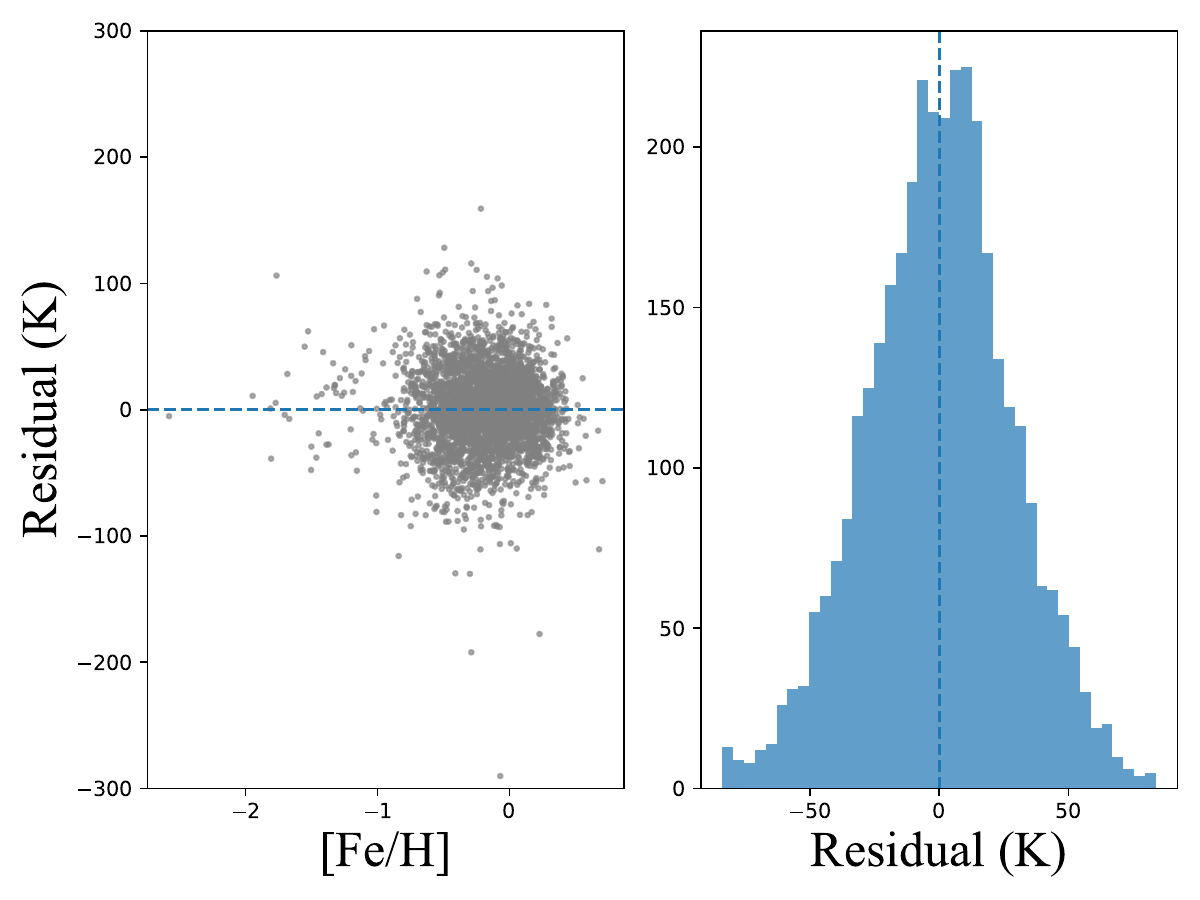}
\includegraphics[width=0.33\textwidth]{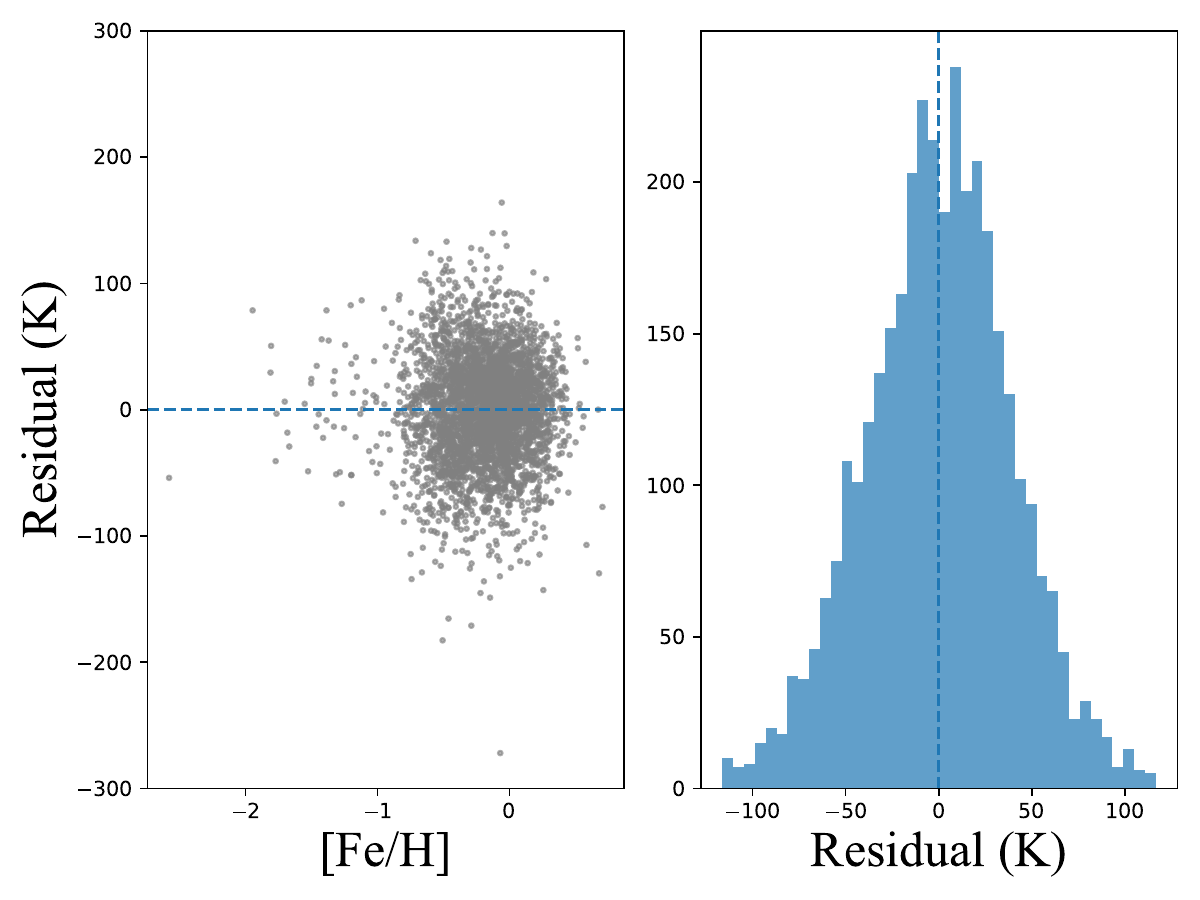}
\includegraphics[width=0.33\textwidth]{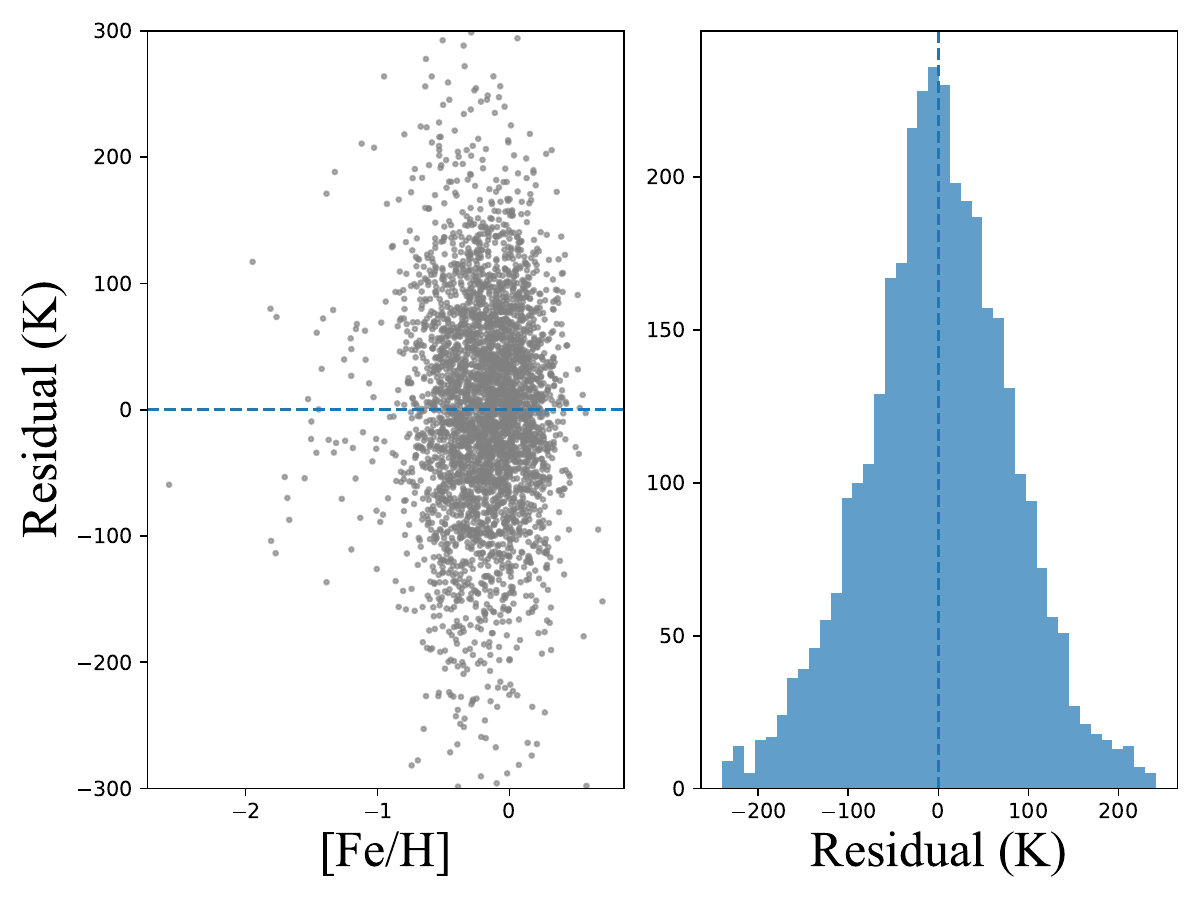}\\

\includegraphics[width=0.33\textwidth]{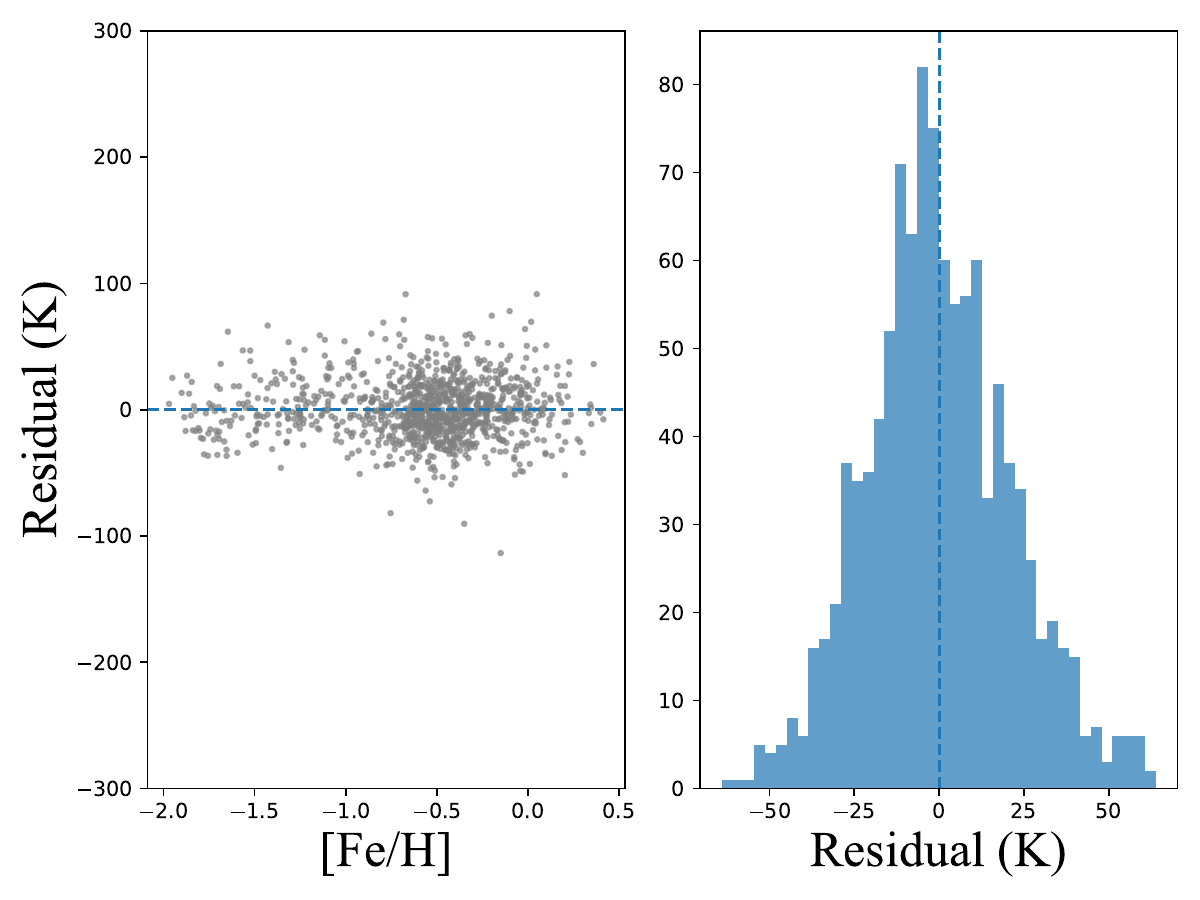}
\includegraphics[width=0.33\textwidth]{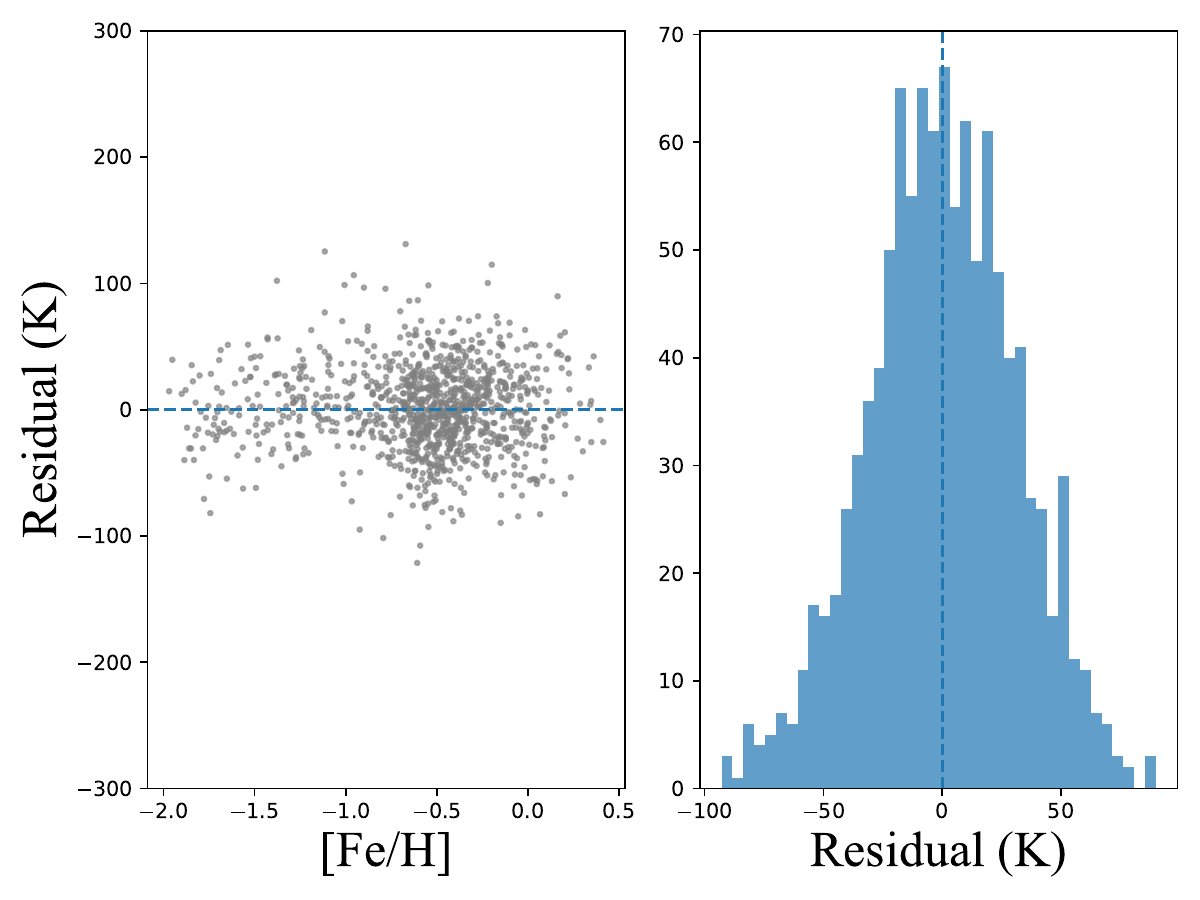}
\includegraphics[width=0.33\textwidth]{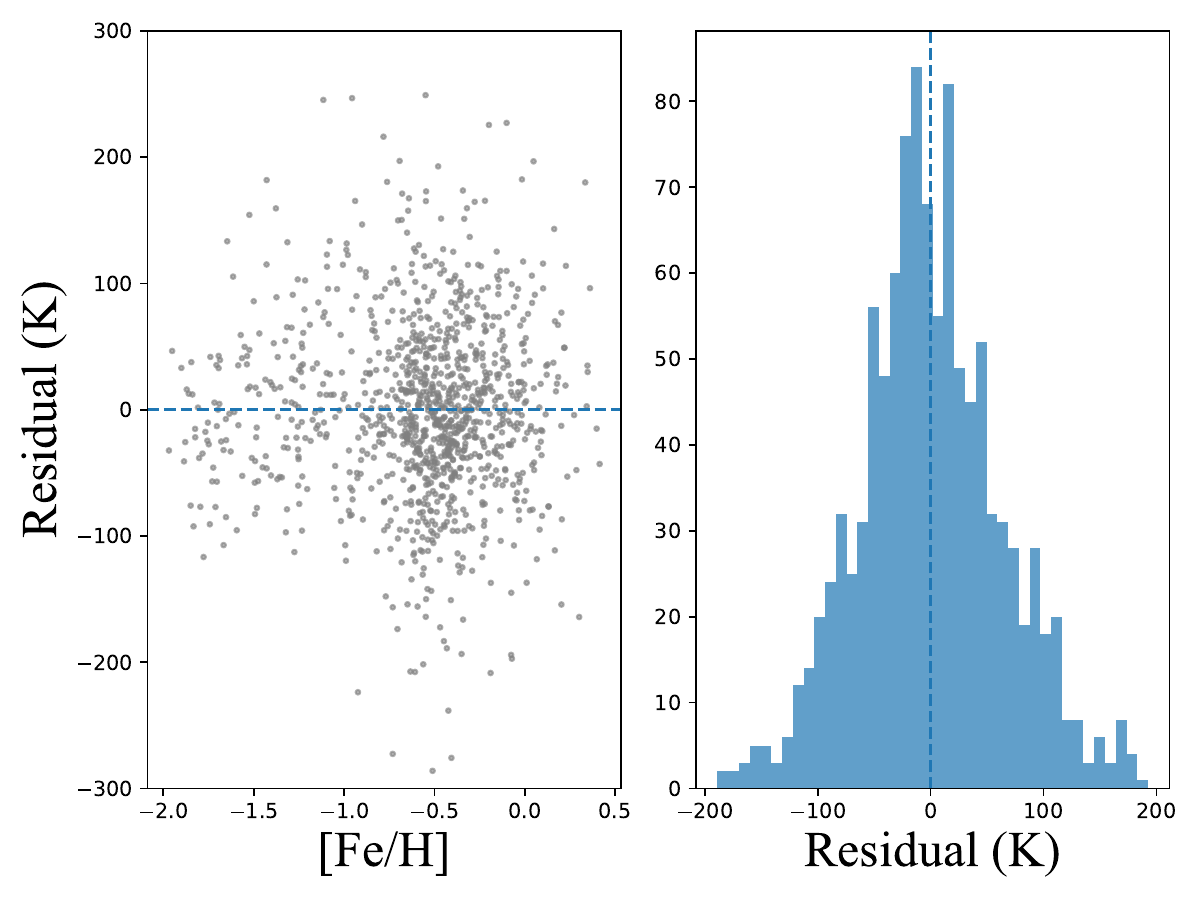}

\caption{Residual diagnostics for dwarfs (top row) and giants (bottom row). Each column corresponds to a colour index: $(g-K_{\rm s})_0$ (left), $(g-J)_0$ (centre), and $(g-r)_0$ (right), representative of long-baseline, intermediate, and purely optical regimes. 
Scatter plots show residuals ($\Delta T_{\rm eff}$) as a function of metallicity, with their distribution shown by histograms.
}
\label{fig:residual_diagnostics}
\end{figure*}

Residual diagnostics for representative colour indices are shown in Figure~\ref{fig:residual_diagnostics}. The residuals are generally centred around zero and display no significant systematic trends with metallicity, indicating that the adopted polynomial forms adequately capture the dominant metallicity dependence of the colour--$\teff$ relations. The residual distributions are approximately Gaussian, with scatter increasing toward shorter wavelength baselines.

Long-baseline colours, particularly $(g-K_{\rm s})_0$ and $(g-J)_0$, exhibit the smallest dispersion and the weakest sensitivity to metallicity. In contrast, purely optical colours show larger scatter, especially at low temperatures and high metallicities, where line blanketing effects become more important.

Including metallicity terms significantly improves the calibration, especially for blue optical colours such as $(u-g)_0$ and $(g-r)_0$. For these indices, the inclusion of metallicity-dependent terms reduces the RMS scatter by up to several tens of Kelvin relative to pure colour relations, demonstrating that metallicity is required to achieve accurate temperature estimates from optical photometry alone.




\subsection{Consistency of dwarfs vs. giants calibrations}

Stars located within the dwarf--giant transition-region are excluded from the calibration procedure and used to assess the consistency between the two sets of relations. Table~\ref{tab:dwarf_giant} compares the temperatures predicted by the dwarf and giant calibrations for these transition-region stars.

Long-baseline colours, such as $(g-K_{\rm s})_0$, $(g-H)_0$, and $(g-J)_0$, show the best agreement between the two calibrations, with mean offsets of only $\sim$15--20~K and dispersions below $\sim$20~K. Intermediate and purely optical colours display progressively larger offsets and scatter, reflecting their stronger sensitivity not only to to metallicity, but also to surface gravity.

In all cases, the dwarf relations predict slightly higher temperatures than the giant relations. This behaviour is physically expected, since decreasing surface gravity lowers atmospheric pressure and modifies line blanketing and continuum opacity, producing systematic colour differences between dwarfs and giants at fixed $\teff$, with the effect varying across different colour indices. The presence of an offset is also a natural consequence of the choice adopted in this work to treat dwarfs and giants as two distinct calibration regimes separated by a sharp boundary in the $\teff$--$\log g$ plane. The observed offsets nevertheless remain comparable to the intrinsic uncertainties of the calibrations. The results indicate that long-baseline optical--infrared colours provide the most stable temperature scale across the dwarf--giant transition and are therefore the most robust indicators within the present calibration framework.

\begin{table}
\centering
\small
\caption{Mean difference and dispersion between effective temperatures predicted by the dwarf and giant calibrations for stars in the transition-region. }
\begin{tabular}{lcc}
\hline
\hline
Colour & Mean ($T_{\rm eff}^{\rm dwarf} - T_{\rm eff}^{\rm giant}$) (K) & $\sigma$ (K) \\
\hline
$(u-g)_0$ & 20.4 & 24.5 \\
$(g-r)_0$ & 45.8 & 45.9 \\
$(r-i)_0$ & 55.1 & 75.3 \\
$(i-z)_0$ & 75.3 &  104.7 \\
$(z-J)_0$ & 50.9 & 67.4 \\
$(J-H)_0$ & 63.4 & 79.7 \\
$(g-i)_0$ & 35.7 & 36.1 \\
$(g-z)_0$ & 28.1 & 29.73 \\
$(g-J)_0$ & 21.4 & 22.7 \\
$(g-H)_0$ & 14.9 & 16.8 \\
$(g-K_{\rm s})_0$ & 14.7 & 16.4 \\
\hline
\end{tabular}
\label{tab:dwarf_giant}
\end{table}

\section{Discussion and Conclusions}\label{sec:concl}

\begin{figure}
    \centering
    \includegraphics[width=0.98\columnwidth]{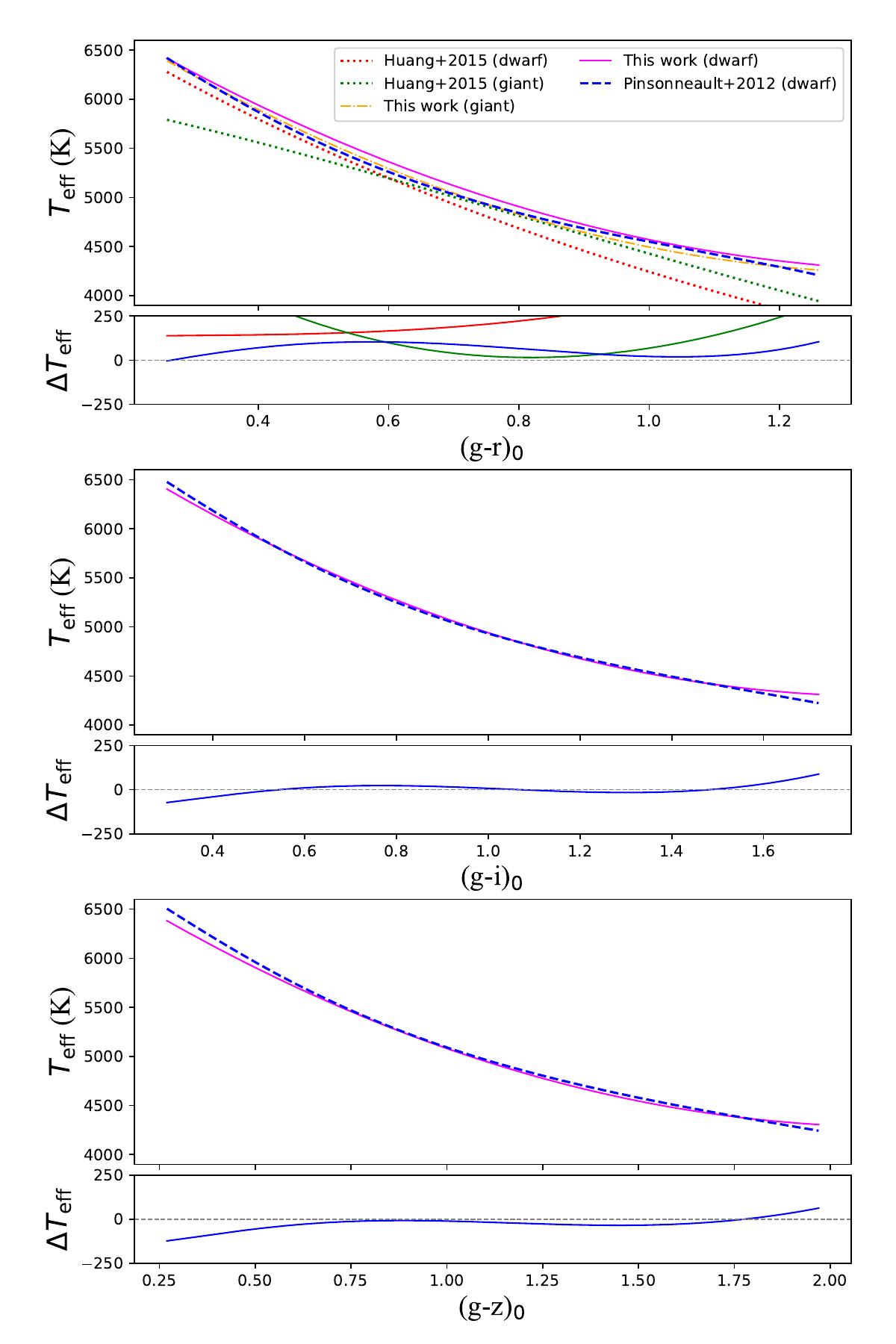}
    \caption{Comparison of the colour--$\teff$ relations derived in this work with the empirical calibrations of \citet{Pinsonneault2012ApJS} and \citet{Huang2015MNRAS} in the SDSS system. For each colour, the top panel shows the predicted effective temperatures as a function of dereddened colour for $(g-r)_0$ (with this colour of \citet{Huang2015MNRAS} added as well), $(g-i)_0$, and $(g-z)_0$, evaluated at a fixed metallicity of $\mathrm{[Fe/H]} = 0.0$. The bottom panel displays the residuals in the sense (this work $-$ Pinsonneault or Huang).
    }
    \label{fig:Pinsonneault2012_comparison}
\end{figure}

Figure~\ref{fig:Pinsonneault2012_comparison} compares our calibrations with those of \citet{Pinsonneault2012ApJS} and \citet{Huang2015MNRAS} using SDSS colours only. Overall, good agreement is found with \citet{Pinsonneault2012ApJS}, with differences typically within $\sim$50~K, whereas offsets easily exceeding $100$~K are appear in the $(g-r)_0$ calibration of \citet{Huang2015MNRAS}. 

\begin{figure*}
    \centering
    \includegraphics[width=0.45\textwidth]{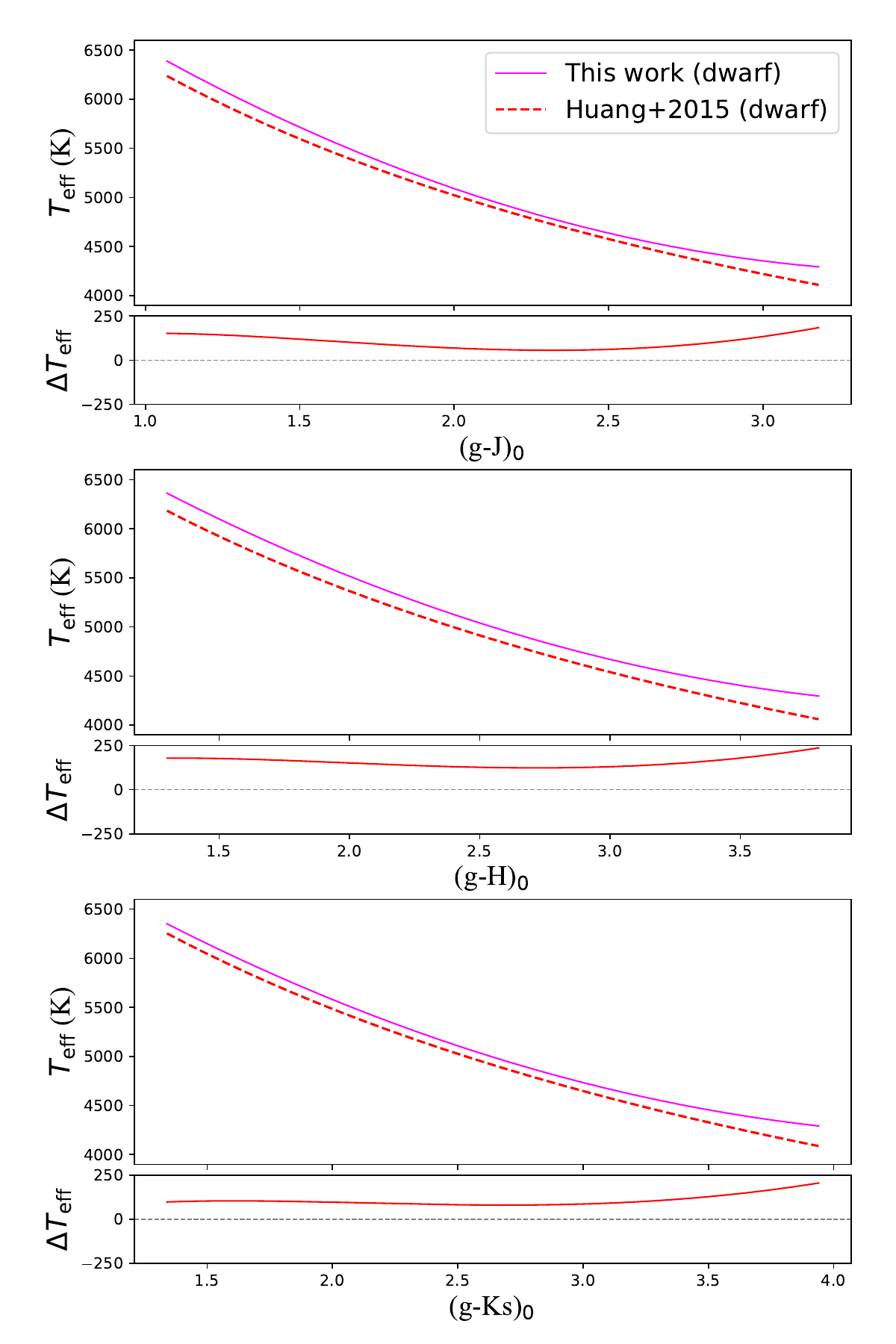}
    \includegraphics[width=0.45\textwidth]{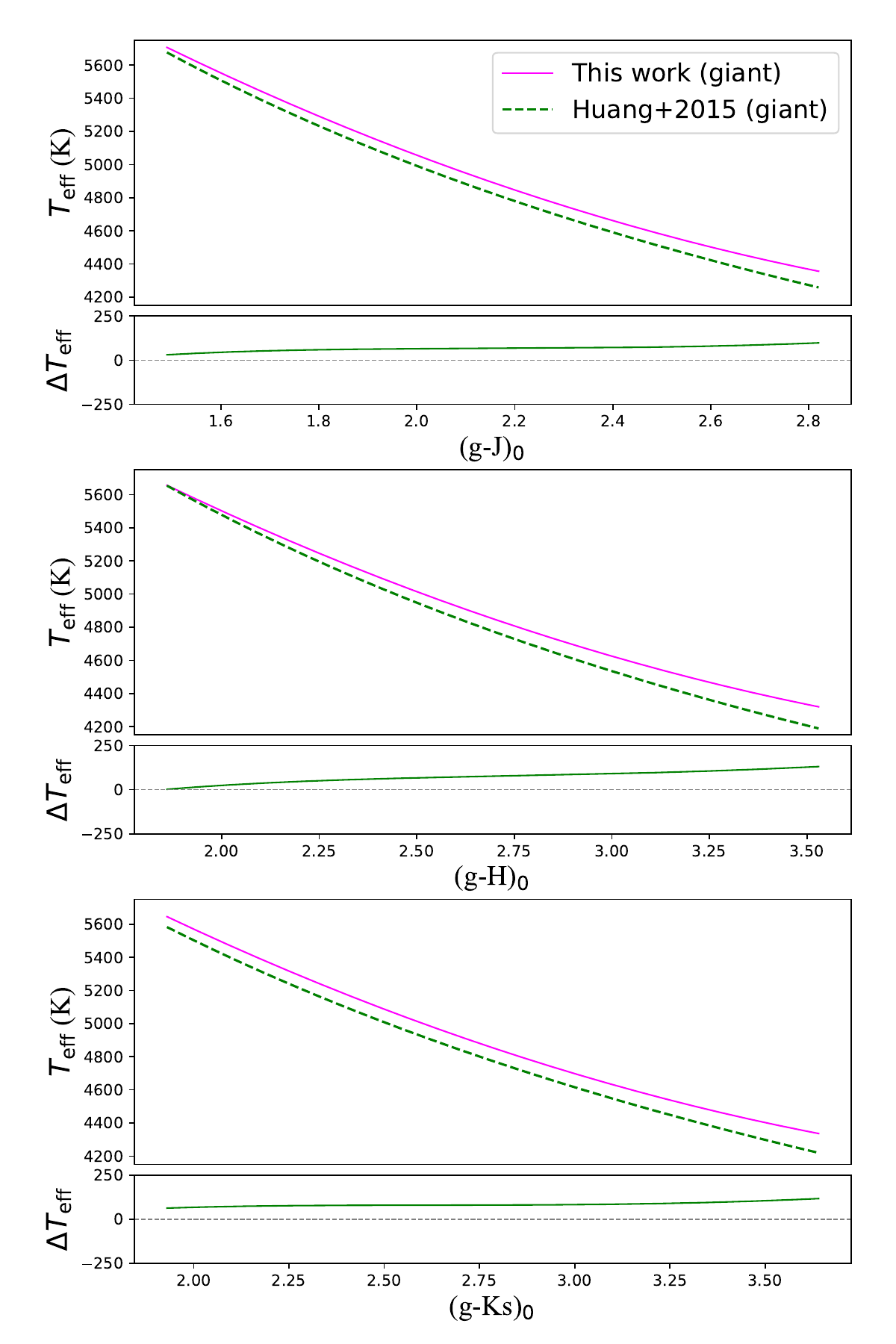}
    \caption{Comparison between the colour--$T_{\rm eff}$ relations derived in this work and the metallicity-dependent calibrations of \citet{Huang2015MNRAS}. The left-hand panels are for dwarfs and the right-hand panels for giants, respectively. The relations are shown for long-baseline optical--infrared colours $(g-J)_0$, $(g-H)_0$, and $(g-K_{\rm s})_0$, evaluated at $\mathrm{[Fe/H]} = 0.0$. Residuals are shown in the lower panels.
}
    \label{fig:Huang2012_comparison}
\end{figure*}

Figure~\ref{fig:Huang2012_comparison} further compares our calibrations against those of \citet{Huang2015MNRAS} using mixed SDSS and 2MASS colour indices. Our relations are systematically hotter by $\sim$50--100~K, particularly at lower temperatures where discrepancies can reach several hundreds of K. This offset is likely related to differences in the adopted temperature scales. In particular, part of the interferometric sample used by \citet{Huang2015MNRAS} relied on earlier angular diameter measurements that were shown to systematically overestimate stellar diameters, leading to cooler effective temperatures \citep{C14}. Additional differences likely arise because the bright stars used by \citet{Huang2015MNRAS} lacked direct SDSS and 2MASS photometry, requiring colour transformations from the Johnson system, which introduce further systematic uncertainties.

In summary, this work provides new empirical colour--$\teff$ relations in the SDSS and 2MASS photometric systems using a homogeneous IRFM temperature scale based on GALAH and APOGEE stars. The calibrations cover FGK dwarfs and giants over the approximate range $4000 \lesssim \teff \lesssim 7000$~K and include explicit metallicity terms which are relevant particularly in optical colours.

A key aspect of the present work is the recalibration of the SDSS $ugriz$ photometric zero-points onto the AB system \citep{zhou26}. Accurate photometric zero-points are particularly important for IRFM temperature determinations, as they directly affect the conversion of observed magnitudes into physical fluxes and therefore propagate into the resulting $\teff$ scale and colour--$\teff$ relations.

The precision of the relations depends strongly on wavelength baseline. Long-baseline optical--infrared colours, especially $(g-K_{\rm s})_0$, $(g-H)_0$, and $(g-J)_0$, provide the tightest and most stable temperature estimates, with typical scatter of $\sim$20--40~K and minimal sensitivity to evolutionary stage. In contrast, purely optical colours exhibit larger dispersion and stronger dependence on metallicity and surface gravity.

Comparison with previous work shows overall good consistency with the \citep{Pinsonneault2012ApJS} SDSS-based calibrations, while systematic offsets relative to \citep{Huang2015MNRAS} are likely driven by differences in temperature scale and photometric transformations. The calibrations presented here provide a homogeneous and internally consistent framework for estimating stellar effective temperatures directly from SDSS and 2MASS photometry, and are well suited for application to large photometric surveys lacking spectroscopy.

\section*{Acknowledgements}

We are grateful to the anonymous referee for the constructive suggestions to strength the presentation. This work is supported by the National Natural Science Foundation of China (NSFC) with grant Nos.12125303, 12288102, 12090040/3, the National Key R\&D Program of China (grant No.2021YFA1600401/ 2021YFA1600403),   the Yunnan Revitalization Talent Support Program—Science \& Technology Champion Project (No.202305AB350003), the New Cornerstone Science Foundation through the XPLORER PRIZE, and the International Centre of Supernovae, Yunnan Key Laboratory (No.202302AN360001). This work is supported by the China Scholarship Council.

This work has made use of data from the European Space Agency (ESA) mission {\it Gaia} (\url{https://www.cosmos.esa.int/gaia}), processed by the {\it Gaia} Data Processing and Analysis Consortium (DPAC, \url{https://www.cosmos.esa.int/web/gaia/dpac/consortium}). Funding for the DPAC has been provided by national institutions, in particular the institutions participating in the {\it Gaia} Multilateral Agreement.
This research has made use of NASA’s Astrophysics Data System, operated by the Smithsonian Astrophysical Observatory under NASA Cooperative Agreement 80NSSC21M0056. It also made use of TOPCAT, an interactive graphical viewer and editor for tabular data \citep{topcat2005ASPC}.

\section*{Data Availability Statement}

The data products are available from the corresponding author upon reasonable request.



\bibliographystyle{mnras}
\bibliography{ref} 




\appendix

\section{Pure colour--$\teff$ relations}\label{appendix}
The colour--$\teff$ relations discussed throughout this paper include a metallicity dependence, as given in Eq.~\ref{eq:Teff_}. For practical applications, however, metallicity information may not always be available, and some colour combinations display both a tight correlation with $\teff$ and only a weak dependence on [Fe/H]. Here we therefore provide fits obtained by removing the [Fe/H] term. It should be noted that while certain colours indeed show little sensitivity to [Fe/H], others retain a significant dependence on it. Consequently, these simplified fits are driven by the bulk metallicity distribution of our sample, which is centred approximately around solar for dwarfs, and subsolar for giants (see Figure \ref{fig:mdf}).

For stars within the dwarf--giant transition region, the same behaviour discussed in Section~\ref{result} remains for the pure-colour relations, with long-baseline colours showing the best agreement between dwarf and giant calibrations. However, both the scatter and systematic offsets are generally larger, as expected from the absence of an explicit metallicity term in the fitting relations.

\begin{figure*}
\centering
\includegraphics[width=0.32\textwidth]{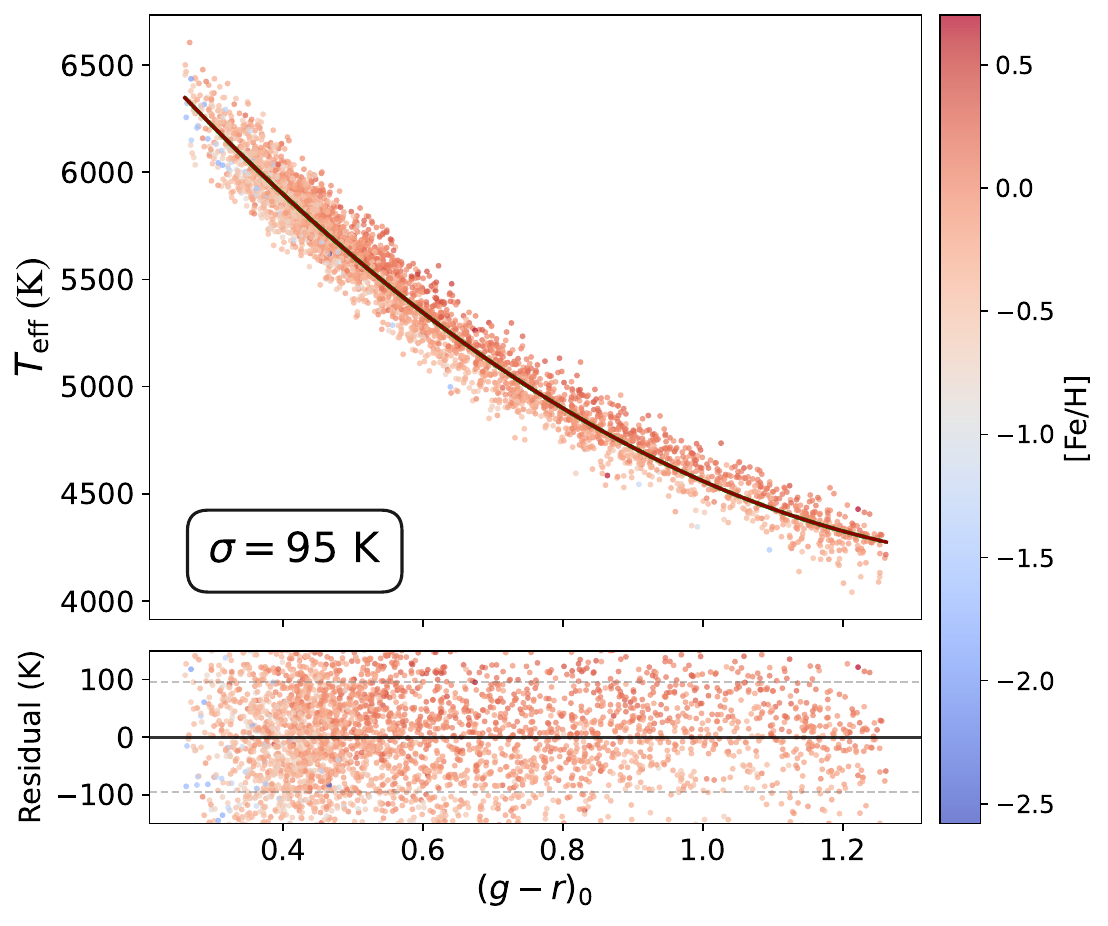}
\includegraphics[width=0.32\textwidth]{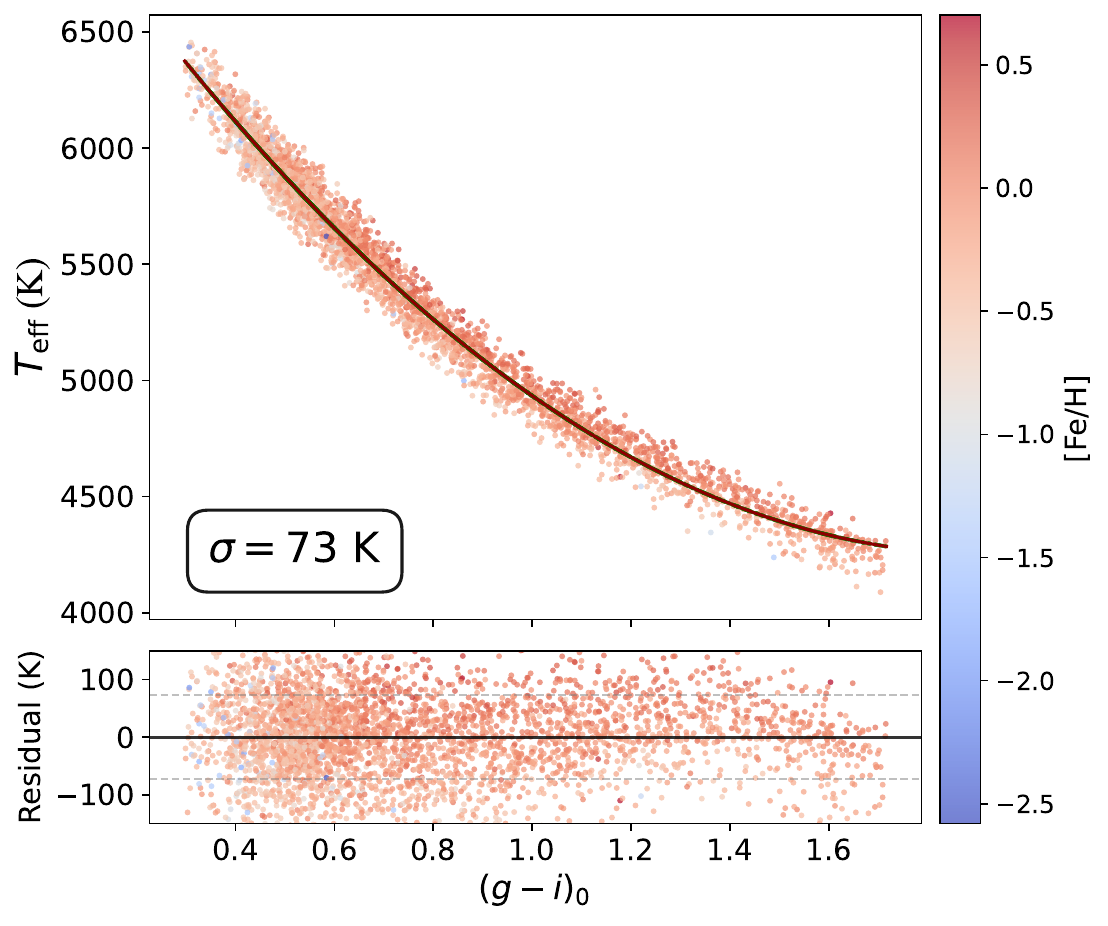}
\includegraphics[width=0.32\textwidth]{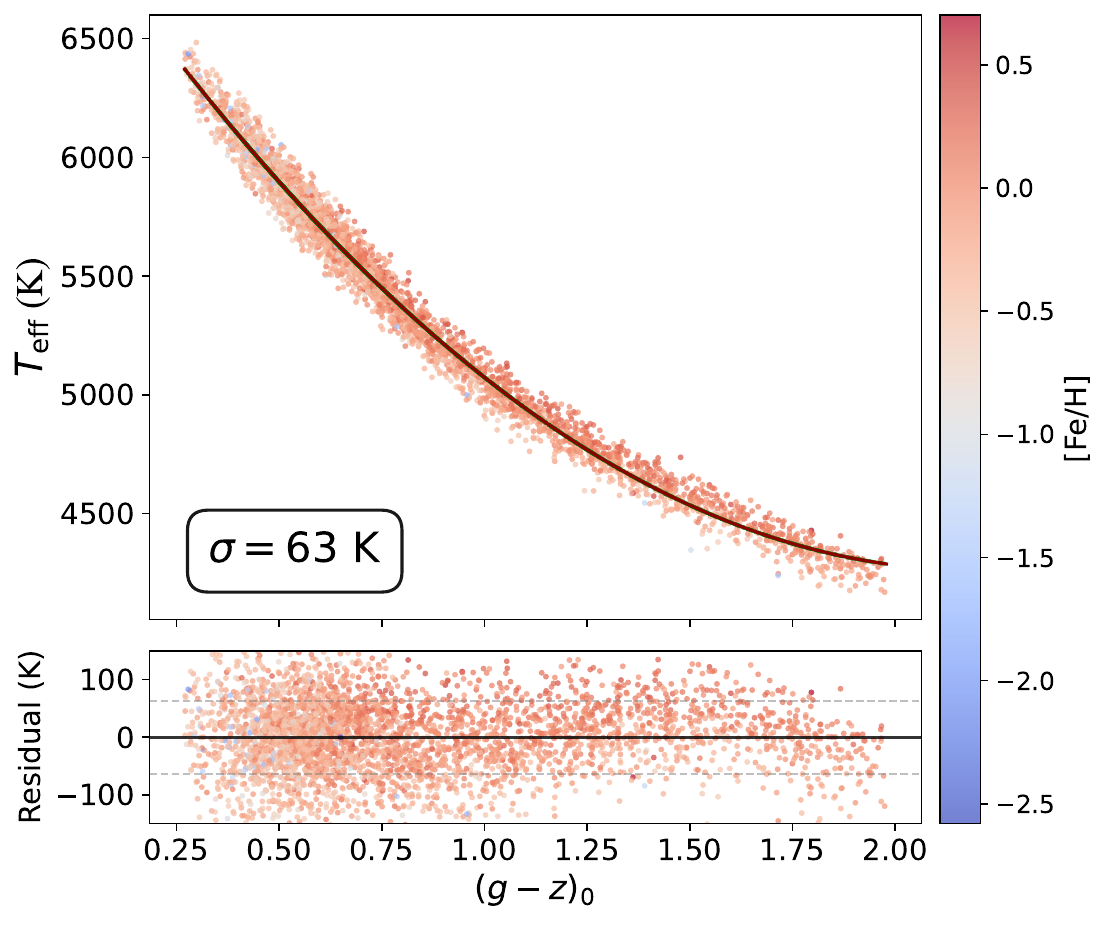}\\

\includegraphics[width=0.32\textwidth]{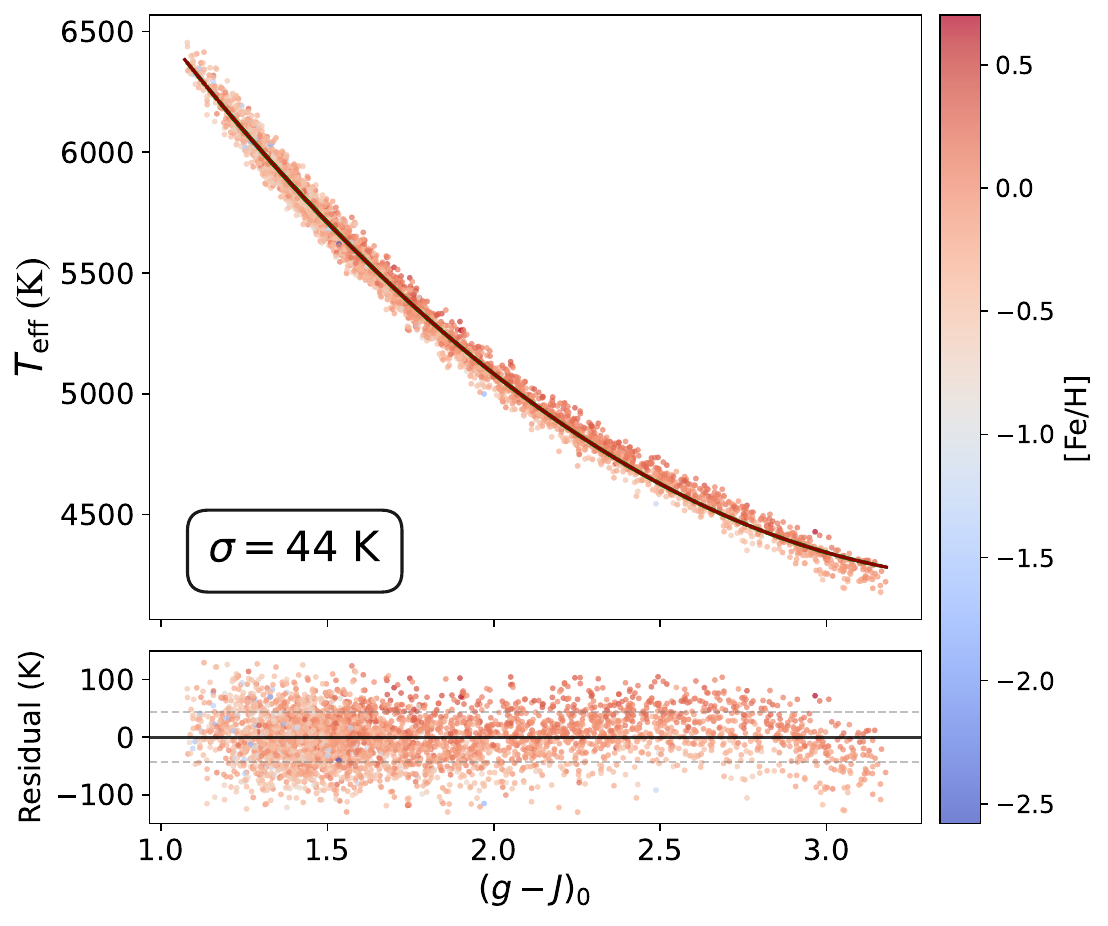}
\includegraphics[width=0.32\textwidth]{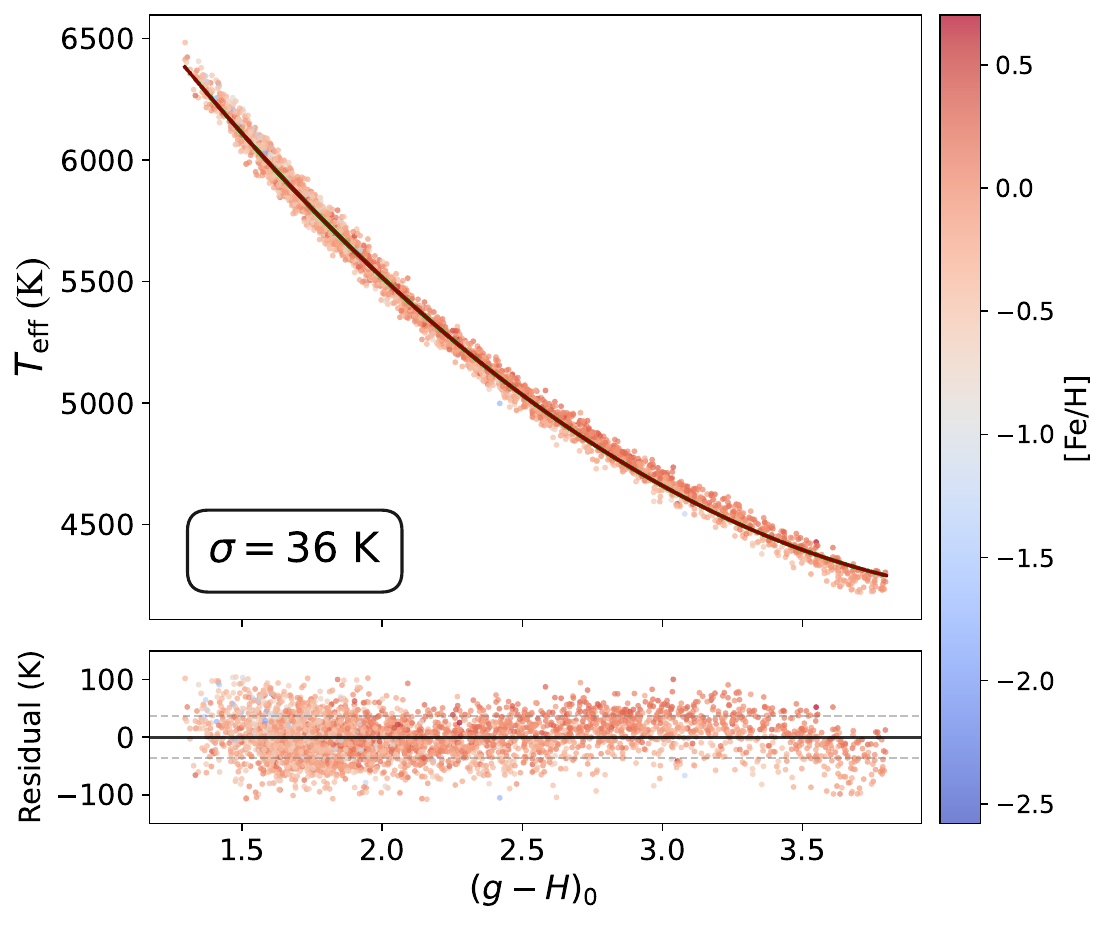}
\includegraphics[width=0.32\textwidth]{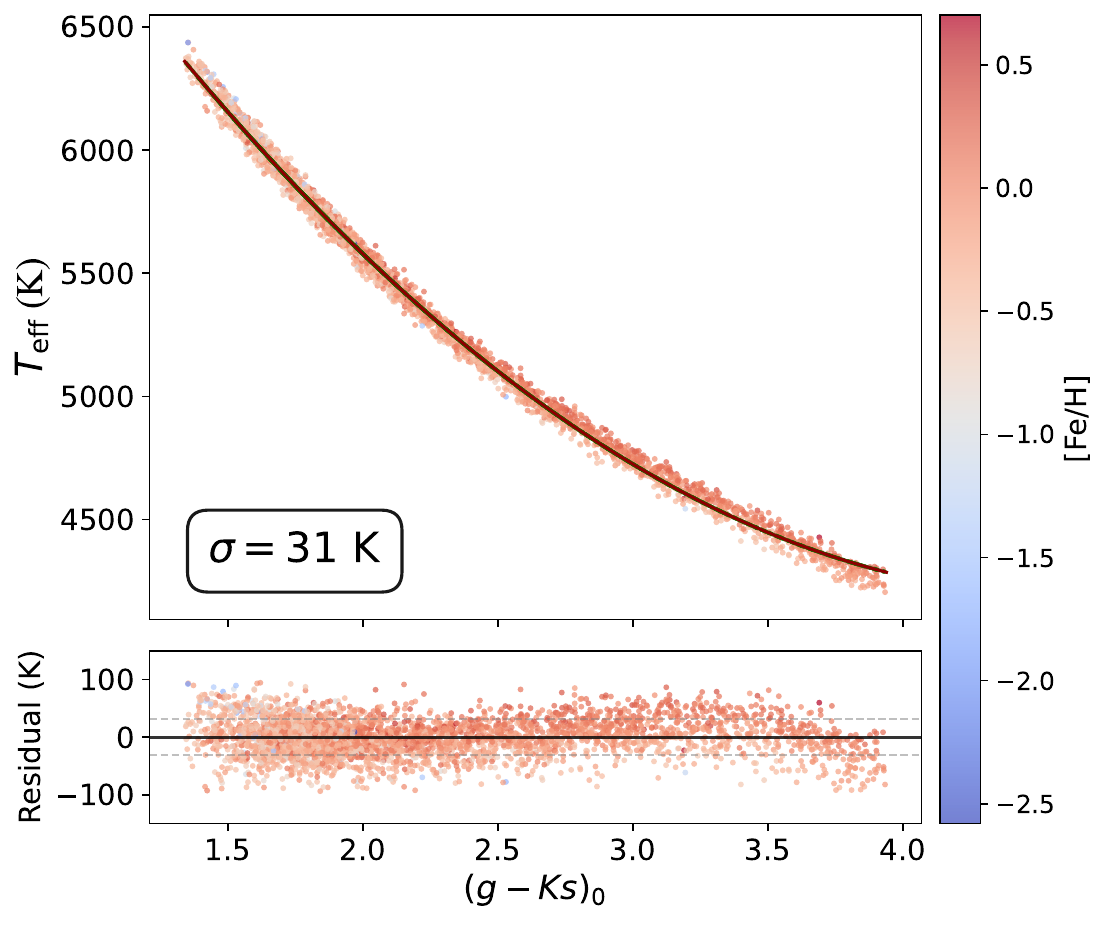}

\caption{Pure colour--$\teff$ relations for dwarf stars. Optical colours (top) are compared with long-baseline optical--infrared colours (bottom). In each plots, continuous lines in upper panels show the fitted relations and lower panels the residuals. The scatter decreases systematically with increasing wavelength baseline.}
\label{fig:pure_relations}
\end{figure*}

\begin{figure*}
\centering
\includegraphics[width=0.32\textwidth]{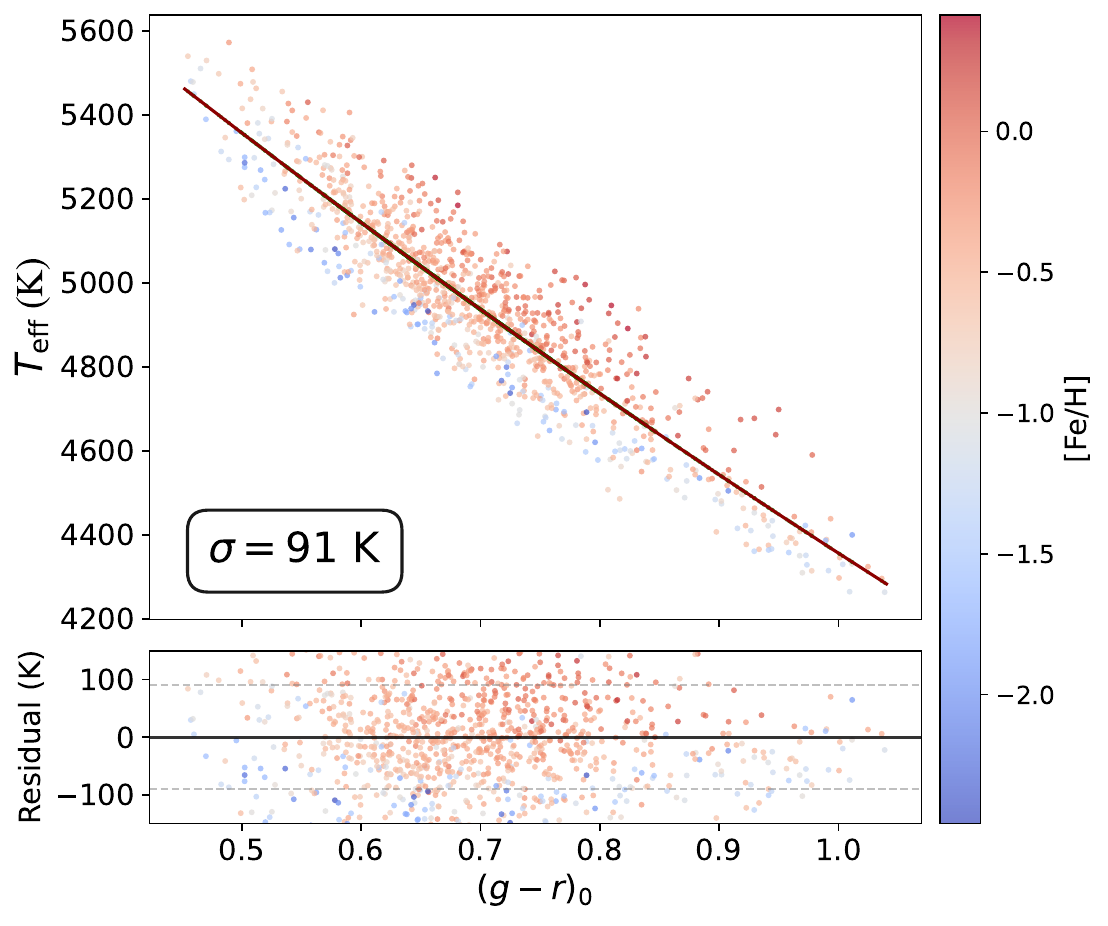}
\includegraphics[width=0.32\textwidth]{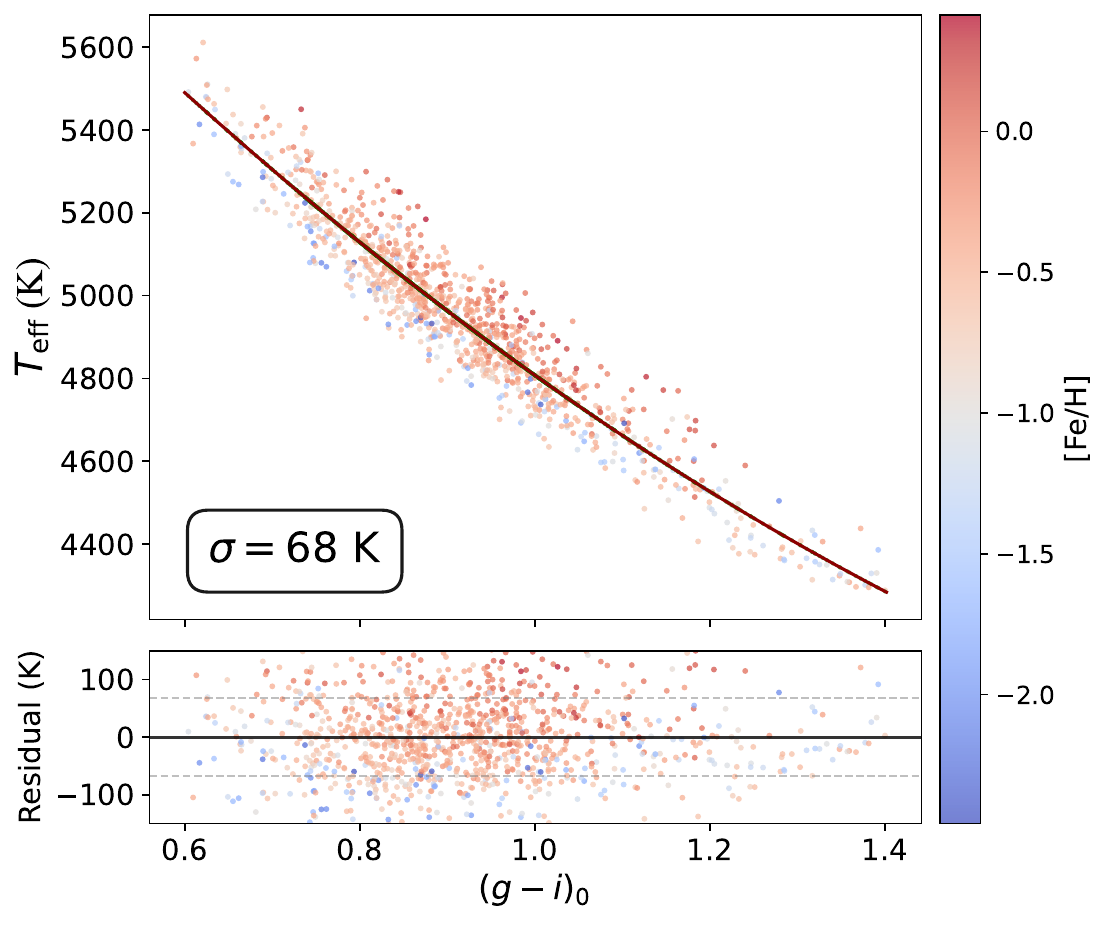}
\includegraphics[width=0.32\textwidth]{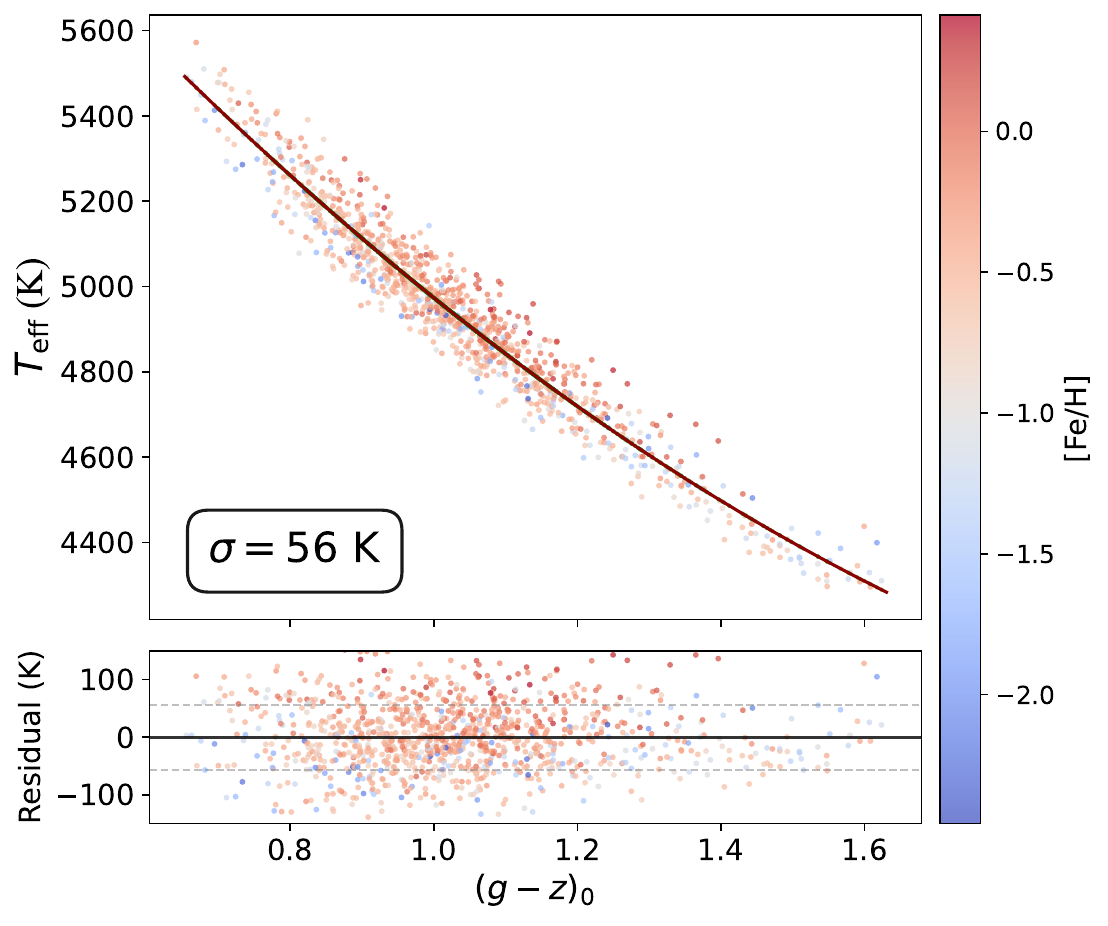}\\
\includegraphics[width=0.32\textwidth]{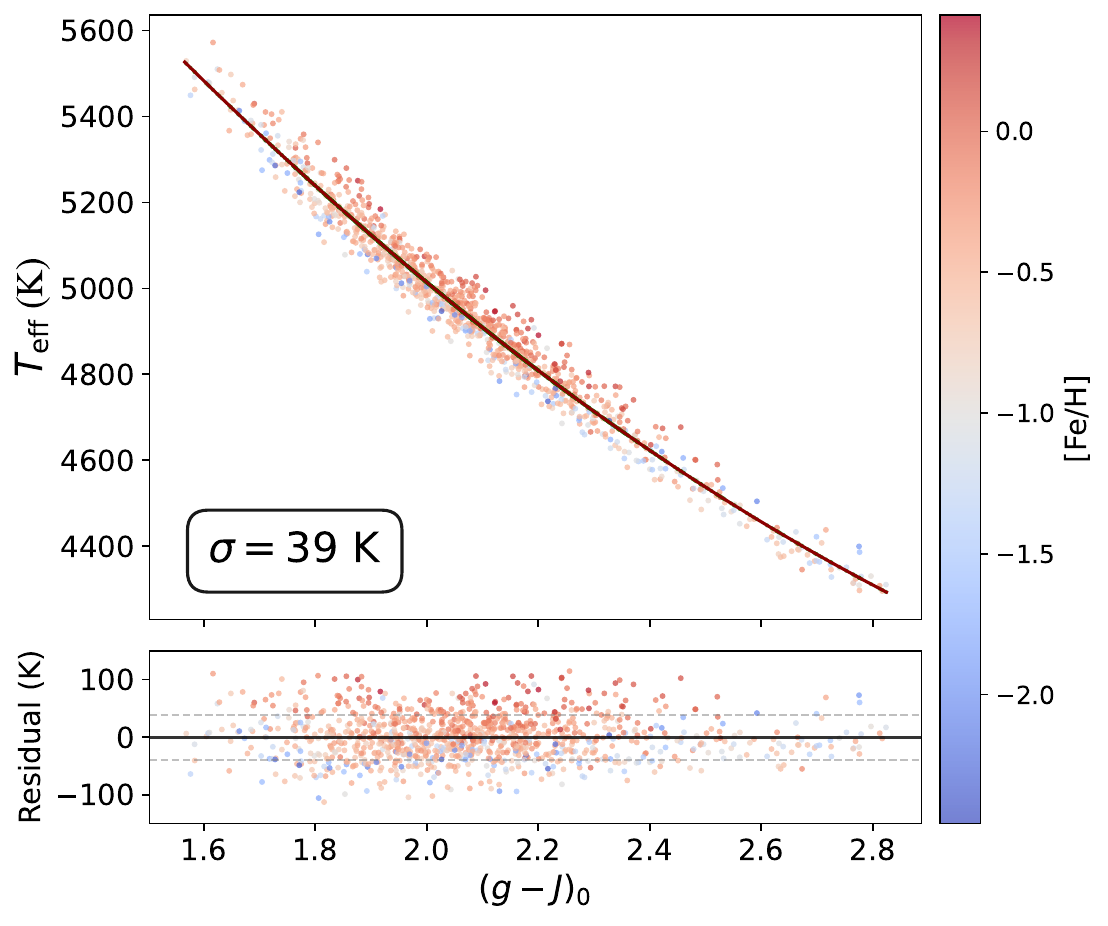}
\includegraphics[width=0.32\textwidth]{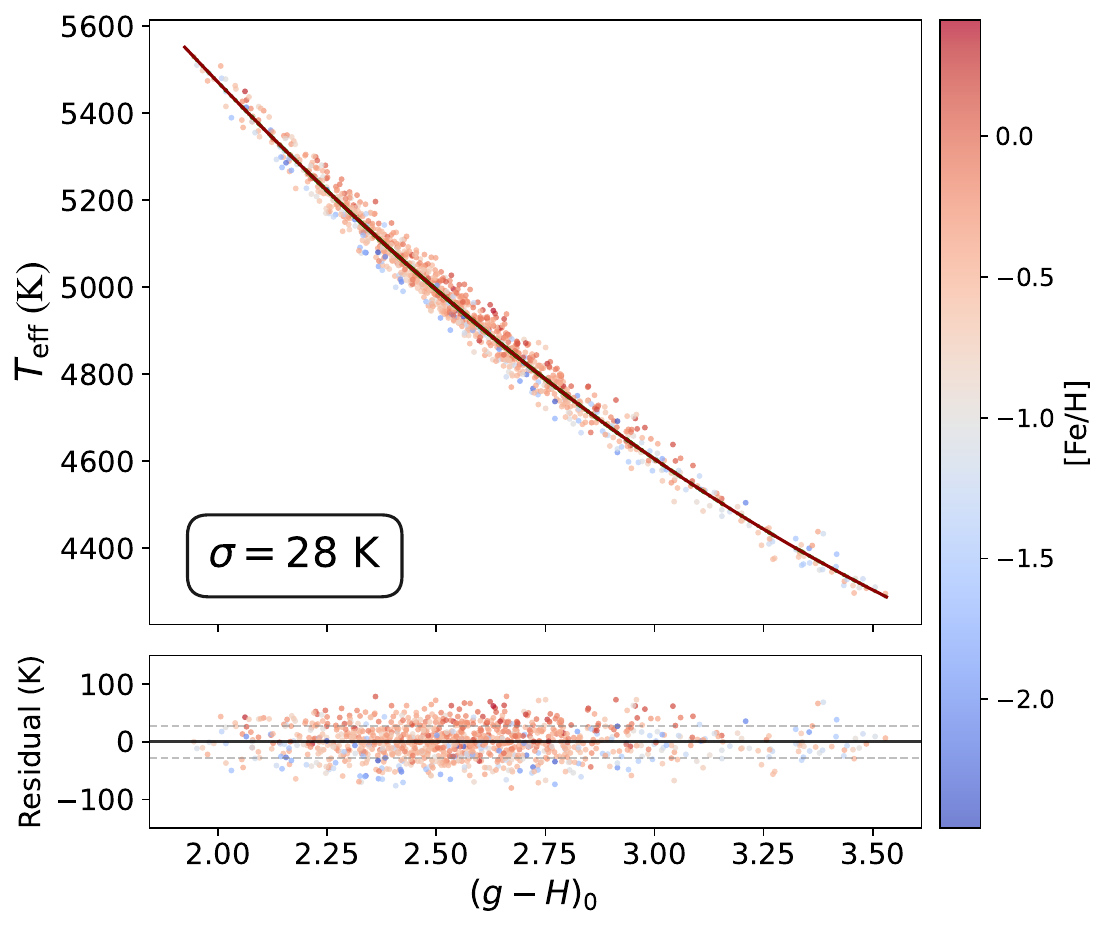}
\includegraphics[width=0.32\textwidth]{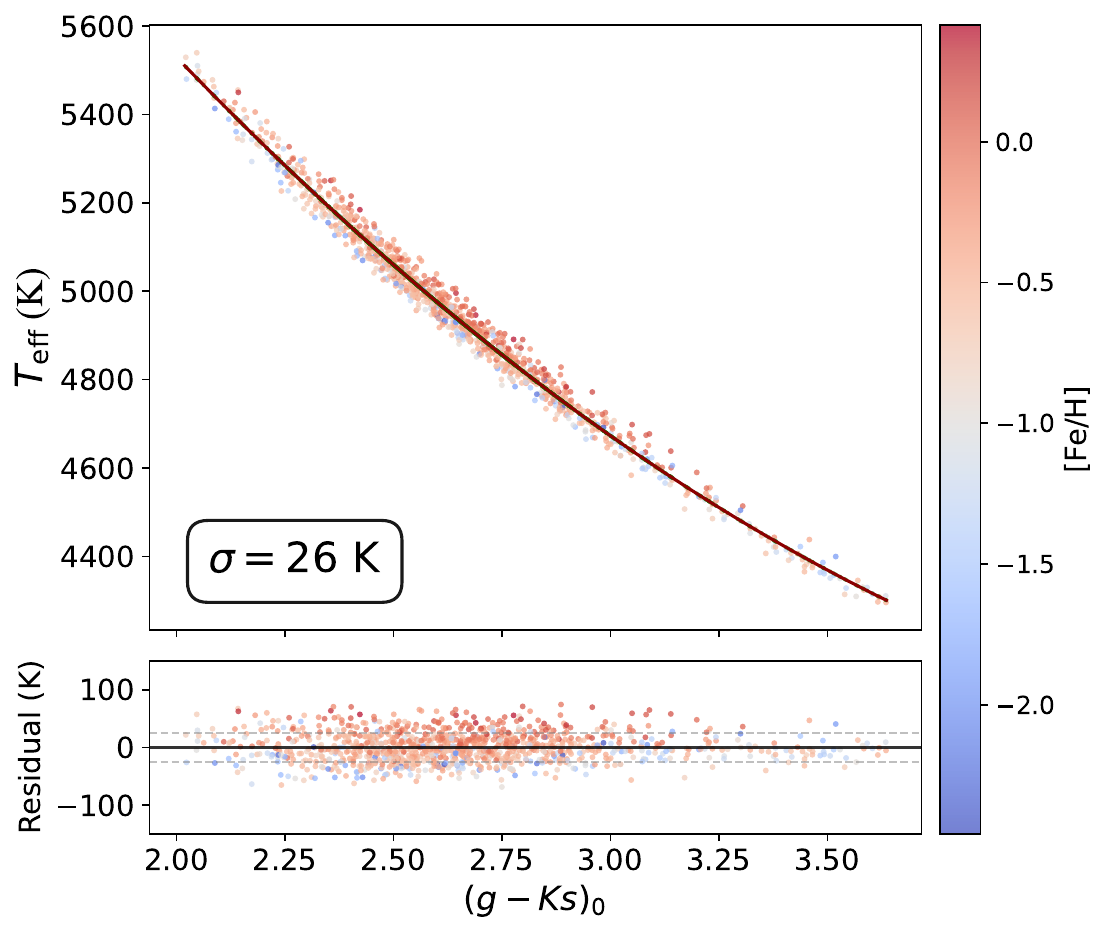}

\caption{Same as Figure~\ref{fig:pure_relations}, but for giant stars.}
\label{fig:pure_relationsGiants}
\end{figure*}

\begin{table*}
\centering
\small
\caption{Polynomial coefficients of pure colour--$T_{\rm eff}$ calibrations for dwarf stars. For each colour index $X$, the fitted relation is:
$\displaystyle
T_{\rm eff} = a_0 + a_1 X + a_2 X^2 + a_3 X^3.
$
}
\label{tab:dwarfs_Purecoeffs}
\begin{tabular}{lcccccccc}
\hline
\hline
Colour & $a_0$ & $a_1$ & $a_2$ & $a_3$ & $N_{\rm total}$ & $N_{\rm used}$ & Colour range & RMS (K) \\
\hline
$(u-g)_0$ & 8032.469 & -2819.947 & 953.675 & -171.157 & 3708 & 3561 & [0.862, 2.506] & 151.1 \\
$(g-r)_0$ & 7317.851 & -4085.065 & 1326.397 & -- & 3708 & 3555 & [0.259, 1.262] & 95.3 \\
$(r-i)_0$ & 6621.580 & -9393.064 & 9068.298 & -- & 3708 & 3523 & [0.022, 0.460] & 149.0 \\
$(i-z)_0$ & 5878.740 & -8368.655 & -6426.688 & -- & 3708 & 3556 & [-0.057, 0.265] & 203.0 \\
$(z-J)_0$ & 11728.200 & -8075.121 & -1530.348 & -- & 3708 & 3561 & [0.784, 1.212] & 142.8 \\
$(J-H)_0$& 7225.116 & -4662.360 & -- & -- & 3708 & 3573 & [0.201, 0.651] & 160.0 \\
$(g-i)_0$ & 7223.107 & -3096.440 & 806.995 & -- & 3708 & 3543 & [0.297, 1.717] & 73.0 \\
$(g-z)_0$ & 7006.269 & -2504.362 & 571.142 & -- & 3708 & 3553 & [0.270, 1.979] & 63.4 \\
$(g-J)_0$ & 8618.482 & -2453.462 & 342.691 & -- & 3708 & 3573 & [1.071, 3.180] & 43.7 \\
$(g-H)_0$ & 8542.380 & -1950.565 & 218.829 & -- & 3708 & 3553 & [1.295, 3.799] & 36.0 \\
$(g-K_{\rm s})_0$ & 8480.684 & -1849.369 & 199.191 & -- & 3708 & 3552 & [1.341, 3.939] & 31.5 \\
\hline
\end{tabular}
\end{table*}

\begin{table*}
\centering
\small
\caption{Same as Table~\ref{tab:dwarfs_Purecoeffs}, but for giants.}
\label{tab:giants_Purecoeffs}
\begin{tabular}{lcccccccc}
\hline
\hline
Colour & $a_0$ & $a_1$ & $a_2$ & $a_3$ & $N_{\rm total}$ & $N_{\rm used}$ & Colour range & RMS (K) \\
\hline
$(u-g)_0$ & 6094.385 & -889.826 & 216.703 & -53.829 & 1129 & 1087 & [1.044, 2.547] & 139.7 \\
$(g-r)_0$ & 6526.237 & -2506.234 & 336.043 & -- & 1129 & 1079 & [0.452, 1.040] & 90.8 \\
$(r-i)_0$ & 5968.543 & -5109.848 & 1861.748 & -- & 1129 & 1087 & [0.116, 0.377] & 124.5 \\
$(i-z)_0$ & 5403.688 & -3363.906 & -5546.307 & -- & 1129 & 1082 & [0.021, 0.233] & 143.6 \\
$(z-J)_0$ & 6932.020 & -294.079 & -1529.247 & -- & 1129 & 1087 & [0.908, 1.222] & 97.9 \\
$(J-H)_0$& 6514.924 & -3169.094 & -- & -- & 1129 & 1086 & [0.331, 0.700] & 108.5 \\
$(g-i)_0$ & 6812.326 & -2505.878 & 500.926 & -- & 1129 & 1082 & [0.600, 1.402] & 68.2 \\
$(g-z)_0$ & 6736.955 & -2174.043 & 410.498 & -- & 1129 & 1080 & [0.653, 1.630] & 56.3 \\
$(g-J)_0$ & 8136.911 & -2047.854 & 243.198 & -- & 1129 & 1084 & [1.566, 2.824] & 39.0 \\
$(g-H)_0$ & 8257.135 & -1744.932 & 175.838 & -- & 1129 & 1086 & [1.923, 3.531] & 28.1 \\
$(g-K_{\rm s})_0$ & 8237.282 & -1685.334 & 165.744 & -- & 1129 & 1080 & [2.019, 3.636] & 25.6 \\
\hline
\end{tabular}
\end{table*}

\begin{table}
\centering
\small
\caption{Same as Table~\ref{tab:dwarf_giant}, but using the pure colour--$\teff$ relations presented in this Appendix.}
\begin{tabular}{lcc}
\hline
\hline
Colour & Mean ($T_{\rm eff}^{\rm dwarf} - T_{\rm eff}^{\rm giant}$) (K) & $\sigma$ (K) \\
\hline
$(u-g)_0$ & 47.4 & 49.1 \\
$(g-r)_0$ & 36.7 & 76.1 \\
$(r-i)_0$ & 6.4 & 40.1 \\
$(i-z)_0$ & -15.2 &  93.6 \\
$(z-J)_0$ & -8.0 & 34.0 \\
$(J-H)_0$ & -3.5 & 27.2 \\
$(g-i)_0$ & 29.1 & 48.0 \\
$(g-z)_0$ & 18.5 & 40.2 \\
$(g-J)_0$ & 10.7 & 27.2 \\
$(g-H)_0$ & 6.6 & 15.8 \\
$(g-K_{\rm s})_0$ & 6.6 & 17.2 \\
\hline
\end{tabular}
\label{tab:Puredwarf_giant}
\end{table}



\bsp	
\label{lastpage}
\end{document}